\documentclass[preprint,journal]{vgtc}       % preprint (journal style)
\ieeedoi{10.1109/TVCG.2019.2934332}

\ifpdf%                                % if we use pdflatex
  \pdfoutput=1\relax                   % create PDFs from pdfLaTeX
  \usepackage{graphicx}                % allow us to embed graphics files
  \DeclareGraphicsExtensions{.pdf,.png,.jpg,.jpeg} % for pdflatex we expect .pdf, .png, or .jpg files
\else%                                 % else we use pure latex
  \usepackage{graphicx}                % allow us to embed graphics files
  \DeclareGraphicsExtensions{.eps}     % for pure latex we expect eps files
\fi%

\graphicspath{{figures/}{pictures/}{images/}{./}} % where to search for the images

\usepackage{microtype}                 % use micro-typography (slightly more compact, better to read)
\PassOptionsToPackage{warn}{textcomp}  % to address font issues with \textrightarrow
\usepackage{textcomp}                  % use better special symbols
\usepackage{mathptmx}                  % use matching math font
\usepackage{times}                     % we use Times as the main font
\usepackage{cite}                      % needed to automatically sort the references
\usepackage{tabu}                      % only used for the table example
\usepackage{booktabs}                  % only used for the table example
\usepackage{multirow}
\usepackage{chngpage}
\usepackage{color, soul}
\usepackage{diagbox}
\usepackage{float}
\usepackage{setspace}
\usepackage{enumitem}
\usepackage{amsmath}
\usepackage{amsfonts}
\usepackage{xcolor,colortbl}
\newcommand{\cmo}{\textcolor[rgb]{0, 0, 0}}
\newcommand{\zw}{\textcolor[rgb]{1, 0, 0}}
\newcommand{\yu}{\textcolor[rgb]{0, 1, 0}}
\definecolor{LightCyan}{rgb}{0.88,0.88,0.88}
\onlineid{1057}

\vgtccategory{Research}
\vgtcpapertype{algorithm/technique}

\title{LassoNet: Deep Lasso-Selection of 3D Point Clouds}
\author{Chen Zhu-Tian, \textit{Student Member, IEEE}, Wei Zeng, \textit{Member, IEEE}, Zhiguang Yang,\\ Lingyun Yu, Chi-Wing Fu, \textit{Member, IEEE}, and Huamin Qu, \textit{Member, IEEE}}
\authorfooter{
\item
Chen Zhu-Tian and Huamin Qu are with the Hong Kong University of Science and Technology. E-mail: zhutian.chen@connect.ust.hk and huamin@cse.ust.hk.

\item
Wei Zeng and Zhiguang Yang are with Shenzhen Institutes of Advanced Technology, Chinese Academy of Sciences. Wei Zeng is the corresponding author. E-mail: \{wei.zeng, zg.yang\}@siat.ac.cn.

\item
Lingyun Yu is with University of Groningen. E-mail: lingyun.yu@rug.nl.

\item
Chi-Wing Fu is with the Chinese University of Hong Kong. E-mail: cwfu@cse.cuhk.edu.hk.
}

\shortauthortitle{Biv \MakeLowercase{\textit{et al.}}: Global Illumination for Fun and Profit}
\abstract{Selection is a fundamental task in exploratory analysis and visualization of 3D point clouds.
Prior researches on selection methods were developed mainly based on heuristics such as local point density, thus limiting their applicability in general data.
Specific challenges root in the great variabilities implied by point clouds (e.g., dense vs. sparse), viewpoint (e.g., occluded vs. non-occluded), and lasso (e.g., small vs. large).
In this work, we introduce LassoNet, a new deep neural network for lasso selection of 3D point clouds, attempting to learn a latent mapping from viewpoint and lasso to point cloud regions.
To achieve this, we couple user-target points with viewpoint and lasso information through 3D coordinate transform and naive selection, and improve the method scalability via an intention filtering and farthest point sampling.
A hierarchical network is trained using a dataset with over 30K lasso-selection records on two different point cloud data.
We conduct a formal user study to compare LassoNet with two state-of-the-art lasso-selection methods.
The evaluations confirm that our approach improves the selection effectiveness and efficiency across different combinations of 3D point clouds, viewpoints, and lasso selections. \emph{Project Website: \textcolor[rgb]{0.10, 0.63, 0.73}{https://lassonet.github.io}}
} % end of abstract


\CCScatlist{ % not used in journal version
 \CCScat{K.6.1}{Management of Computing and Information Systems}%
{Project and People Management}{Life Cycle};
 \CCScat{K.7.m}{The Computing Profession}{Miscellaneous}{Ethics}
}

\teaser{
  \centering
  \includegraphics[width=0.97\linewidth]{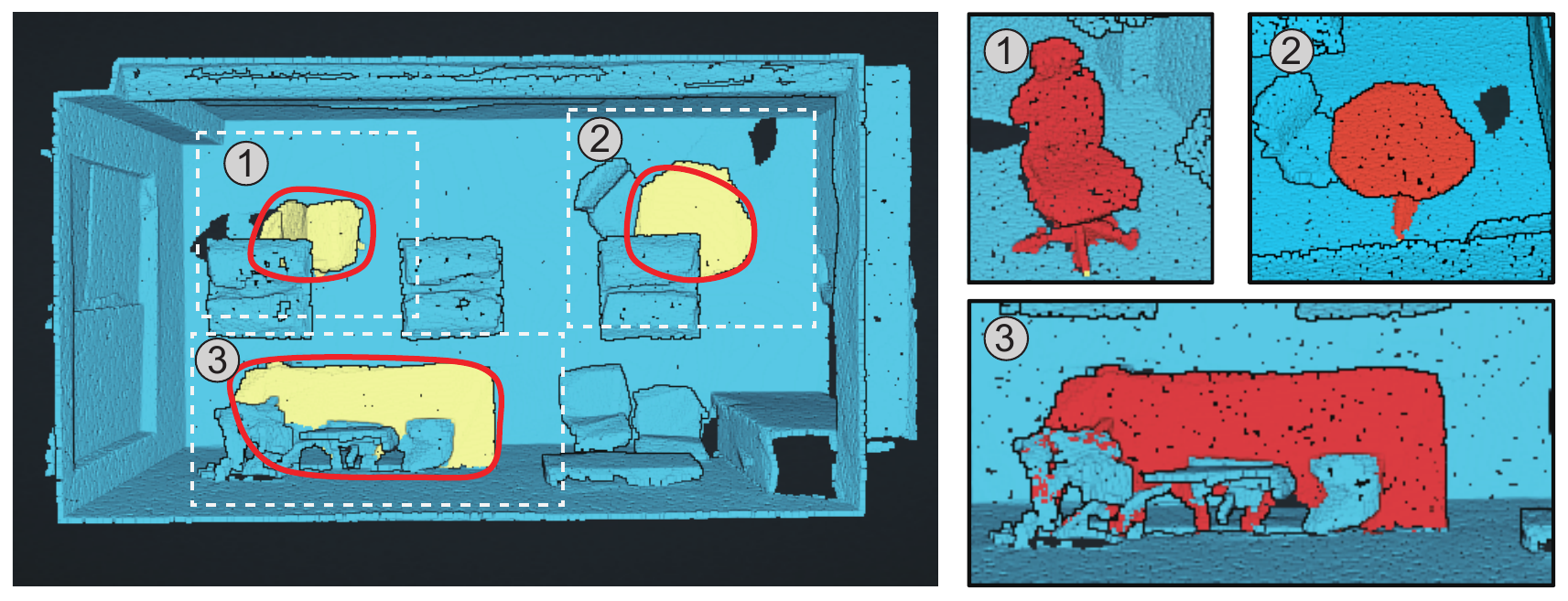}
  \vspace{-4mm}
  \caption{
  LassoNet enables effective lasso-selection of 3D point clouds based on a latent mapping from viewpoint and lasso to target point clouds.
  LassoNet is particularly efficient for selecting multiple regions \cmo{(insets 1, 2, 3)} in a complex scene \cmo{(left)}, since no viewpoint changing is required to select occluded points.
  \cmo{Notice here the insets are viewed from different viewpoints.}
  }
	\label{fig:teaser}
}

\vgtcinsertpkg

\begin{document}


%% Introduction + related works ==> 2 pages
%% Problem statement + methods ==> 2 pages
%% Datasets ==> 0.5 page
%% Model experiments ==> 1.5 page
%% User study ==> 1.5 page
%% discussion + conclusion ==> 1.5 page
%% total 9 pages
\firstsection{Introduction}

\maketitle

Vast amounts of 3D point clouds have been collected from various sources, such as LiDAR scanning and particle simulation.
% Laser scanner and depth sensing technologies have captured vast amounts of 3D point clouds.
Exploratory analysis and visualization of point clouds show benefits in many applications,
including astronomy, autonomous navigation, and scene reconstruction.
% To enable interactive visualization, a key functionality is selection~\cite{wills_1996_selection}.
Selection is a fundamental task in exploratory analysis of point clouds.
However, designing effective selection for 3D point clouds is challenging, 
especially when the visualization is projected onto a planar 2D surface~\cite{keefe_2013_reimagining}.
The challenge comes from several perspectives:
%% raw data is lack of semanitcs labels/information YZG
1) \textit{data}: a point cloud \cmo{often} consists of a set of unlabeled points, \emph{i.e.}, no information of what label each point holds; %~\cite{bacim_2014_slice};
2) \textit{visualization}: projecting points in 3D space to 2D surface can easily cause occlusion and visual cluttering; %~\cite{tong_2017_glyphlens};
3) \textit{interaction}: input devices such as mouses and touchscreens operate in a limited 2D space.
% , which has limited abilities to perform 3D operations such as \cmo{rotation}~\cite{chen_1988_study}.
% and 4) \textit{user}: transformation from 3D to 2D can greatly affect a user's perceptual and cognitive processes~\cite{bruckner_2018_model}.

Compared to other selection means of picking and brushing, lasso-selection is considered more appropriate for 3D point clouds~\cite{Yu2016}.
The interactive process derives an appropriate subset of points ($P_s$) from a point cloud dataset ($P$), based on a lasso ($L$) input drawn on a 2D surface and the current viewpoint ($V$).
This work aims to develop an advanced lasso-selection method for exploring 3D point clouds.
We consider that the requirements for such a lasso-selection method include:

\begin{itemize}[leftmargin=*]

% \vspace{1mm}
\item
\textit{Efficient}:
The method \cmo{should enable users to complete point selection as soon as possible,
for which a necessary condition is that 
the computation should be finished in a short time}.
% and provides feedbacks upon selecting operations at an interactive rate.
% This requires short computation time.

% \vspace{1mm}
\item
\textit{Effective}:
The derived subset of points $P_s$ is expected to match with the target points of a user's intention ($P_t$).
Here, we employ Jaccard distance ($d_J$) to measure difference between $P_s$ and $P_t$.

% \vspace{1mm}
\item
\textit{Robust}:
The method should be robust to variability implied by data (e.g., dense \emph{vs} sparse), viewpoint (e.g., occluded \emph{vs} non-occluded), and lasso drawing (e.g., small \emph{vs} big).

\end{itemize}

Recently, a number of lasso-selection methods for point clouds have been proposed ~\cite{Yu2012, shan_2012_interactive, Yu2016}.
The methods can be categorized as \textit{structure-aware} that depend on characteristics of point clouds, e.g., local point density~\cite{Yu2012, shan_2012_interactive}, or \textit{context-aware} that further take into account the lasso location and shape~\cite{Yu2016}.
All methods employ a heuristic that users intend to select regions of higher local point density, which is valid for astronomical datasets where users are typically interested in galaxy cores~\cite{Yu2012, Yu2016} or halos~\cite{shan_2012_interactive}.
However, many other point clouds do not exhibit this property.
In this case, these density-based selection techniques lose their advantages and often select unintended clusters that have higher densities of points than target clusters.

%Moreover, these methods can be intrinsically considered as a \textit{segmentation} process that depends on features of both point clouds and user inputs: a bounding box covering the selected area is split into equal-volume cubes, which are then segmented into selected or unselected according to the density fields. Points inside the same cube can by no means be separated, which we can pose negative effects on selection accuracy.

In this work, we approach lasso-selection of 3D point clouds from a new angle: lasso-selection is regarded as a latent mapping function ($f$) between point clouds ($P$), viewpoint ($V$), and lasso ($L$), \emph{i.e.}, $f(P, V, L) \rightarrow P_s$.
We hereby introduce LassoNet, a learning-based approach to seek an optimal mapping based on deep learning techniques.
LassoNet integrates deliberated modules to tackle challenges of:
(i) \textit{data heterogeneity} induced by 3D point clouds, viewpoint, and 2D lasso $-$ we associate them using 3D coordinate transformation and naive selection (Sec.~\ref{ssec:inter_encode}); 
(ii) \textit{generalizability} caused by \cmo{widely ranging number of points and varying point densities} $-$ we employ an \cmo{intention filtering} and farthest point sampling (FPS) algorithm (Sec.~\ref{ssec:filter_sample}).
We build a hierarchical neural network to learn local and global features (Sec.~\ref{ssec:net}) from a dataset consisting of over 30K lasso-selection records on two different point cloud corpora.
Model experiments show that LassoNet can effectively learn the mapping (Sec.~\ref{sec:model}).
We also conduct a formal user study showing that LassoNet advances state-of-the-art lasso-selection methods (Sec.~\ref{sec:study}).
% . in effectiveness and efficiency over various conditions (Sec.~\ref{sec:study}).
% Qualitative user feedback also recommend our approach due to its simplicity and robustness.

\vspace{1mm}
\noindent
Our contributions are as follows

\begin{itemize}[leftmargin=*]

% \vspace{1mm}
	\item
	We develop LassoNet $-$ a deep neural network that models lasso-selection of 3D point clouds as a latent mapping from viewpoint and lasso to point cloud regions.
	To our knowledge, this is a first attempt of \textit{learning-based} approach that successfully addresses limitations of existing \textit{heuristics-based} methods.

% \vspace{1mm}
	\item 
	We address the challenges of data heterogeneity using 3D coordinate transformation and naive selection, and generalizability using \cmo{intention} filtering and farthest point sampling. We further build a hierarchal neural network to improve network performance.
	
% \vspace{1mm}
	\item
	We train LassoNet on a new dataset with over 30K lasso-selection records, and release the code and dataset to enable future studies on lasso-selection of point clouds.
	A formal user study confirms LassoNet fulfills the \textit{efficient}, \textit{effective}, and \textit{robust} requirements.

\end{itemize}

\section{Related Work}
\label{sec:related}

% We group related work in three categories:
% \textit{lasso-selection for 3D point clouds}, introducing existing methods for selecting point clouds in 3D space projected on a 2D surface;
% \textit{deep learning for interaction}, outlining a recent trend of employing deep learning to facilitate interactions in visualization;
% and \textit{deep learning on point clouds}, discussing recent developments of deep learning techniques on point clouds.

% We summarize related work in the following three categories:

%%%%%%%%%%%%%%%%%%%%%%%%%%%%%%%%%%%%%%%
\vspace{1mm}
\noindent
\textbf{Lasso-Selection for 3D Point Clouds.} \
% \label{ssec:3d_sel}
Selection is a fundamental task in interactive visualization~\cite{wills_1996_selection}.
Designing effective interaction theories and methods is regarded as a main challenge for scientific visualization~\cite{keefe_2013_reimagining}.
Systematic reviews of 3D objects selection techniques can be found in~\cite{laviola_2017_user, argelaguet_2013_survey}.
% Bruckner et al.~\cite{bruckner_2018_model} modeled spatial directness of interaction techniques considering spaces of \textit{data}, \textit{visualization}, \textit{interaction}, \textit{output}, \textit{manipulation} and \textit{user}.
% Following the model, we have discussed various factors hindering selection of 3D point clouds in the previous section.
%
Specifically, lasso-selection is preferable for interacting with 3D point clouds projected on a 2D surface~\cite{Yu2016}.
% The great variability implied by data, visualization, and interaction hinders the design of an \textit{efficient}, \textit{effective}, and \textit{robust} lasso-selection for 3D point clouds.
% Next, we briefly summarize existing methods on lasso-selection for 3D point clouds.
%
An ultimate problem here is how to deduce user-intended region in 3D view frustum from a lasso on 2D surface.
% Raycasting that casts a ray from a point is frequently used to \textit{pick} distinct objects~\cite{argelaguet_2009_efficient, chadwick_2005_exploring}.
% However, raycasting techniques usually select the first object it hits, which cannot work when the target object is obscured by other objects.
% Thus, they are often used when there are few targets which can be easily recognized.  
% Casting from points of a lasso extrudes a cone or cylinder in the view frustum.
Cone/Cylinder-selection~\cite{Forsberg:1996:ABS, Steed:2006:GMS} is a basic method, which selects all objects within a geometry (\emph{i.e.}, cone or cylinder) extruded from a lasso.
% since all objects inside of the selecting geometry get selected.
The selections can be refined by moving the cone/cylinder~\cite{Steed2004}.
% Early lasso-selection methods select multiple objects within the cone, and refine the selection by moving the cone~\cite{Steed2004}.
% Haan et al.~\cite{Haan2005} scored objects of user-intension based on the object movements.
% In line with this idea, 
Owada et al.~\cite{Owada2005} improved volume data selection by segmenting the data according to user-drawn stroke.
This idea of deducing regions of user intention based on underlying data characteristics inspired \textit{structure-aware} lasso-selection methods~\cite{Yu2012, shan_2012_interactive}. %assume that users intend to select regions with high density fields.
A series of dedicated methods were developed, including TeddySelection and CloudLasso~\cite{Yu2012}.
WYSIWYP (`what you see is what you pick') technique~\cite{wiebel_2012_wysiwyp} can be integrated to provide instant feedback~\cite{shan_2012_interactive}.
Recently, \textit{context-aware} methods~\cite{Yu2016} that further take into account the lasso position and shape were developed.
% These methods are referred as CAST~\cite{Yu2016}.

\begin{figure}
	\centering
	\includegraphics[width=0.95\columnwidth]{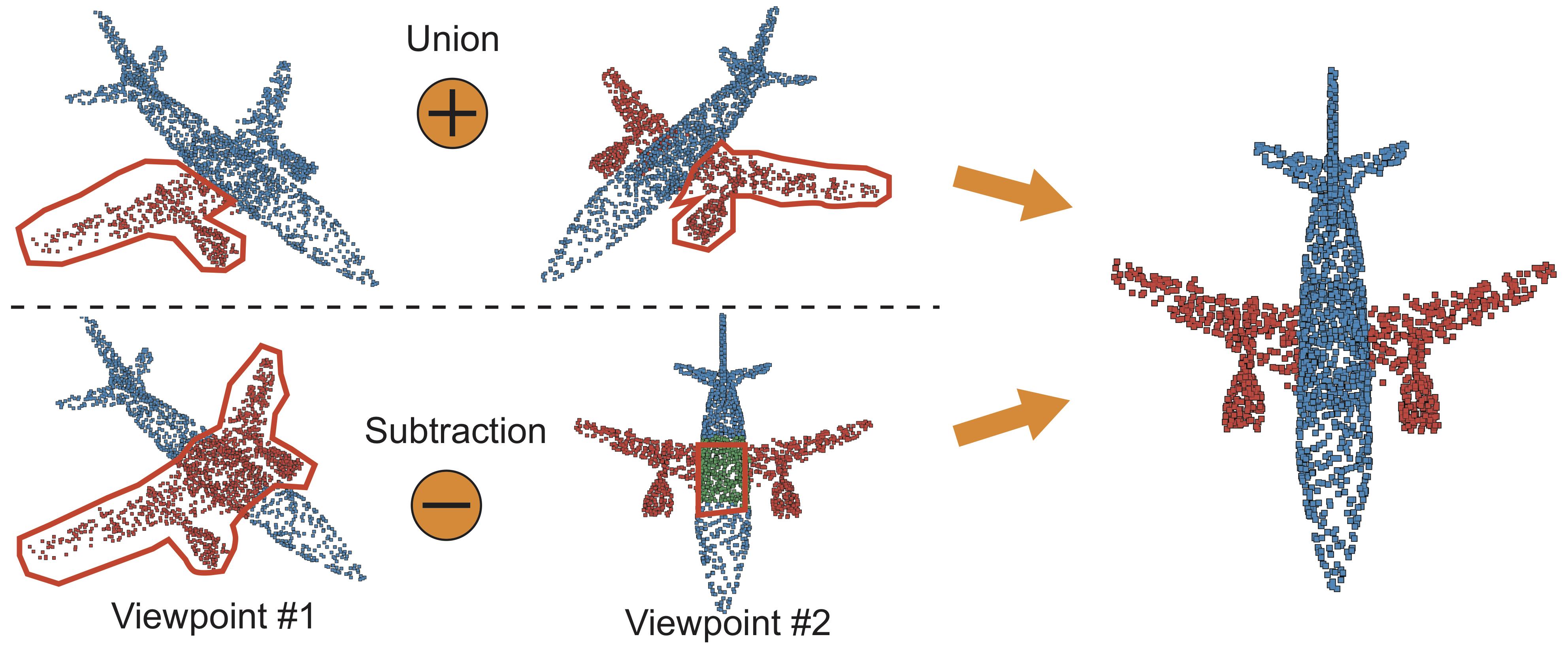}
	\vspace{-2mm}
	\caption{To select both two wings of the airplane, two sequential lassos from different viewpoints are required, and selection results are combined using either \textit{union} (top) or \textit{subtraction} (bottom) Boolean operations.
	}
	\label{fig:lassoselections}
	\vspace{-4mm}
\end{figure}

\vspace{1mm}
\noindent
\textbf{Limitations of Existing Lasso-Selection Techniques}. Conventional CylinderSelection methods select all the points enclosed by the frustum extruded from an input lasso.
In case target points are not located in the same area, Boolean operations of multiple lasso-selections are needed.
For instance, it is a common task to select both the left and right wings of an airplane.
As illustrated in Fig.~\ref{fig:lassoselections}, users can complete the task by either joining the right and left wing regions with \textit{union} (top), or removing the body part region from a larger selection with \textit{subtraction} (bottom).
In a more complicated scenario, e.g., to select yellow target points in a more complex scene as in Fig.~\ref{fig:teaser}, users need to draw multiple lassos from different viewpoints.
Much time will be spent on finding a suitable viewpoint for each target. 
% Moreover, in case a lasso is not precise enough to enclose the target points, unintended points will be selected.
% Thus, for the general lasso selection techniques, a precise lasso from a prefect viewpoint is necessary but difficult to achieve by users. 

CAST~\cite{Yu2016} have been developed for making 3D point cloud selections more intuitive and efficient. 
The main idea of CAST family techniques is to infer a user-intended selection region based on properties of point clouds and lasso drawings. 
% The techniques are particularly effective for point clouds with density variations. 
They employ a series of heuristics based on density information.
 % to make appropriate connections between point clouds and lassos, and finally deduce a 3D volume. 
Therefore, CAST is particularly effective for selections in 3D point cloud datasets with varying point densities, for instance, astronomy simulations of galaxy collisions and N-body problems.
% Moreover, they are appropriate for domain studies which require to select and explore a specific range of data based on varying scalar properties besides density.
However, not all 3D point clouds, e.g., ShapeNet and S3DIS datasets studied in this work (see Sec.~\ref{sec:model}), exhibit this property.
% In these datasets, density does not have an intrinsic meaning, making it inappropriate to be a selection criterion.
Taking the airplane (Fig.~\ref{fig:lassoselections}) extracted from ShapeNet for an example, all parts share almost the same point density.
CAST hereof will perform similarly to CylinderSelectionL: all points within the frustum extruded from a lasso will be selected.

To make lasso-selections more \textit{robust} and \textit{efficient}, we should go beyond scalar properties of point density. 
One possible solution is to add more intrinsic features of point clouds, such as heat kernel signature~\cite{sun_2009_concise, bronstein_2010_scale}.
Nevertheless, the approach would need tremendous trial-and-error processes to find suitable parameters (which may not even exist).
Instead, we opt to \emph{learning-based} approach, as we envision that a deep neural network can effectively capture intrinsic features of point clouds, and eventually learn an optimal mapping $f(P, V, L)$.

\vspace{1mm}
\noindent
\textbf{Deep Learning for Interaction.} \
% \label{ssec:dl_inter}
Recent years have witnessed the burst of deep learning techniques, benefiting many fields such as image process and natural language processing.
The visualization community has also been contributing to deep learning.
On the one hand, many visualization systems have been developed to `\textit{open the black box}' of deep learning models through visual understanding, diagnosis, and refinement~\cite{Liu2017, Hohman2018}.
On another hand, emerging researches have employed deep learning to address domain-specific tasks, such as to classify chart types~\cite{jung_2017_chartsense}, to measure the similarity between scatterplots~\cite{ma_2018_scatternet}, and even to perceive graphical elements in visualization~\cite{haehn_2019_evaluating}.
The community has also conducted several pieces of research on exploiting deep learning techniques to facilitate user interactions.
% Sentence fragment?
Fan and Hauser~\cite{Fan2018} modeled user brushing in 2D scatterplot as an image, which can be handled by a convolutional neural network (CNN) to predict selected points.
The method greatly improves selection accuracy, meanwhile preserves efficiency.
Han et al.~\cite{Han2018} developed a voxel-based CNN framework for processing 3D streamlines, by which clustering and selection of streamlines and stream surfaces are improved.

Inspired by them, we also model lasso-selection of 3D point clouds using deep learning.
However, in our case, point clouds are distributed in 3D space. 
Thus CNNs (e.g.,~\cite{Fan2018}) designed for 2D images are not feasible.
Point clouds datasets exhibit great diversity, \emph{e.g.}, sparse \emph{vs} dense, balanced \emph{vs} imbalanced density.
Voxel-based neural network~\cite{Han2018} that divides the volume into a low resolution of $64^3$ voxelization can dramatically reduce prediction accuracy.
Instead, we employ a feature-based deep neural network (DNN) that has becoming more popular for processing 3D point clouds.

\begin{figure*}[t]
\centering % avoid the use of \begin{center}...\end{center} and use \centering instead (more compact)
\includegraphics[width=1.9\columnwidth]{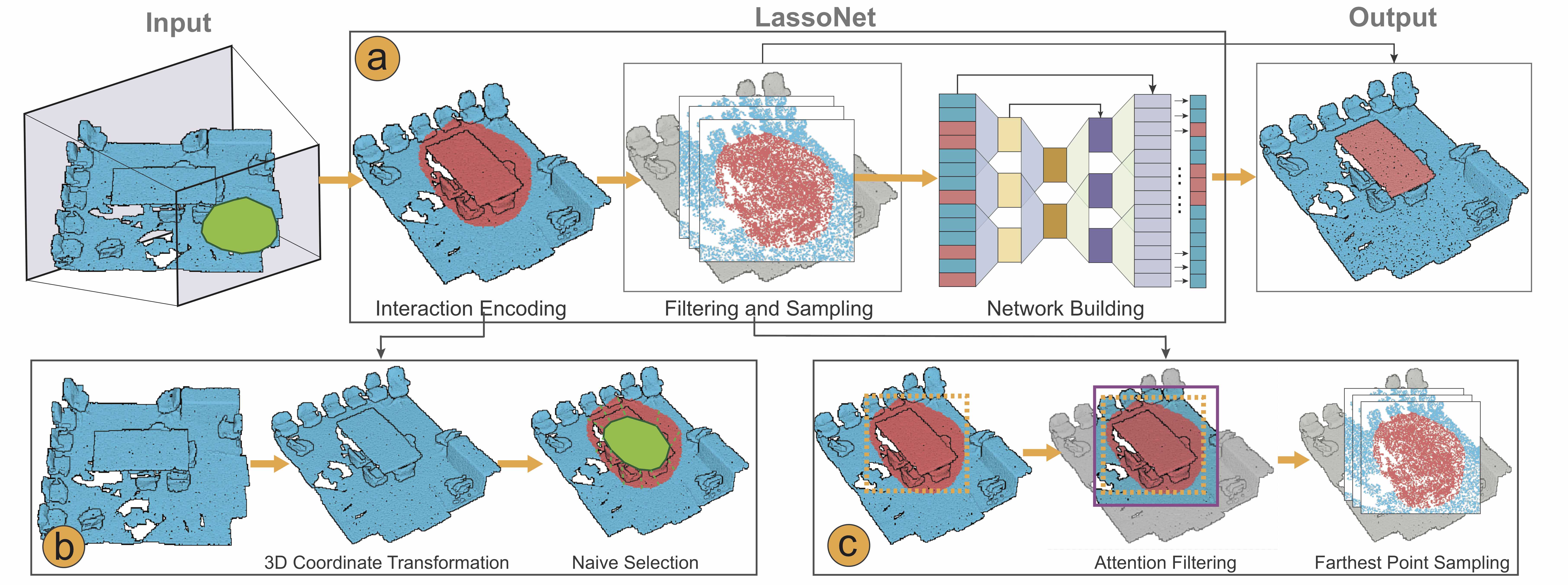}
\vspace{-3mm}
\caption{
LassoNet consists of three stages:
In \textit{Interaction Encoding} stage, we associate point cloud with viewpoint and lasso through 3D coordinate transformation and naive selection;
In \textit{Filtering and Sampling} stage, we reduce the amount of points for network processing through intention filtering and farthest point sampling.
Lastly, we build a hierarchical neural network in \textit{Network Building} stage.
}
\vspace{-3mm}
\label{fig:pipeline}
\end{figure*}

%%%%%%%%%%%%%%%%%%%%%%%%%%%%%%%%%%%%%%%%%%%%%%%%%%%%%%%%
\vspace{1mm}
\noindent
\textbf{Deep Learning for Point Clouds.} \
Point cloud is a special type of 3D geometric data that can be processed by deep learning
 % has gained increasing attention from computer vision and graphics researchers for 3D shape analysis
 ~\cite{wang_2017_cnn}.
Based on how to model 3D shapes into CNN processable units, prior researches can be categorized as:
\textit{Volumetric CNNs} (e.g.,~\cite{Wu2015, Maturana-2015-6018}) convert a 3D shape into voxels and apply a 3D CNN over voxels, which faces a critical problem of sparsity, especially for processing non-uniformly distributed point clouds.
\textit{Multiview CNNs} (e.g.,~\cite{Qi2016, su_2015_multi-view}) render 3D shapes into 2D images from multiple viewpoints, and then apply 2D CNNs to analyze them.
However, in this work viewpoint is a parameter in the latent mapping that we expect the network to learn.
\textit{Spectral CNNs} (e.g.,~\cite{Masci2015, bronstein_2017_geometric}) learn geometric features defined on 3D manifold meshes, which however, are not easy to construct from 3D point clouds.
Many studies (e.g.,~\cite{Fang2015, Guo:2015:MLV:2870647.2835487}) have employed \textit{feature-based DNNs} that convert 3D data into a vector and apply fully connected network to classify the shape.
This approach can be seamlessly integrated with shape features, making it popular for processing 3D shapes now.

However, it is extremely challenging to build a suitable DNN for point clouds, as too many features can be derived from enormous points.
Qi et al.~\cite{Qi2017} successfully addressed the challenge with PointNet, which consists of multiple layers of feature transformation, multi-layer perceptron, and max pooling.
Based on PointNet, they further developed a hierarchical neural network~\cite{Qi2017a} that improves the prediction accuracy.
Recently, a series of follow-up works (e.g.,~\cite{Wang2018b, Shi2018, Ye2018}) were conducted to address more complex tasks.
We also build LassoNet upon Qi's work.
To our knowledge, this is the first extension that has been developed to facilitate user interaction.
We show how to tackle challenges of data heterogeneity and scalability using domain knowledge in visualization and human-computer interaction (Sec.~\ref{sec:lassonet}).

\section{Problem Formulation}
\label{ssec:prob}

The scope of this work is constrained to lasso-selection of 3D point clouds using a 2D input device (e.g., mouse) to draw lassos on a 2D surface (e.g., a desktop monitor).
Other input and display devices, such as 3D hand gestures and virtual reality HMDs, are out of the scope.
% As illustrated in Fig.~\ref{fig:selection}, 
The selection process involves three components:

\begin{itemize}[leftmargin=*]

% \vspace{1.5mm}
\item
\textit{Point cloud ($P$)}: A point cloud $P$ consists of a set of points $\{\mathbf{p}^i_{obj} := (x^i_{obj}, y^i_{obj}, z^i_{obj})\}_{i=1}^n$, where $x^i_{obj}, y^i_{obj}, z^i_{obj}$ indicate position of the point $\mathbf{p}^i_{obj}$ in a 3D \textit{object} space $\mathbb{R}^3_{obj}$.
$n$ is the total number of points, which can be in a wide range from a few thousand (ShapeNet dataset) to hundreds of thousands (S3DIS and astronomical datasets).
Unlike images that are made up of organized pixel arrays, a point cloud holds no specific order of points.
Besides, many point clouds in real-world are unsegmented.
We would like the selection method being applicable to them.
Nevertheless, the unordered and unsegmented properties compound the difficulty of effective lasso-selection.

% \vspace{1.5mm}
\item
\textit{Viewpoint ($V$)}:
A viewpoint $V$ is determined by many factors, including camera position and direction, field of view (FOV), and perspective/orthogonal projection.
When a visualization starts, FOV and projection type are usually fixed.
Users can control the viewpoint by translating and rotating the camera.
Before drawing a lasso on the screen, users tend to find an optimal viewpoint that minimizes occlusion for the target points.

% \vspace{1.5mm}
\item
\textit{Lasso ($L$)}:
A lasso $L$ can be represented as an ordered list of points $\{\mathbf{l}^i_{scn} := (u^i_{scn}, v^i_{scn})\}_{i=1}^m$, where $u^i_{scn}, y^i_{scn}$ indicate position of a point $\mathbf{l}^i_{scn}$ in a 2D \textit{screen} space $\mathbb{R}^2_{scn}$.
The lasso $L$ meets two requirements:
1) \textit{Closed}:
In case the user-drawn stroke is not closed, we enclose it by connecting its start ($\mathbf{l}^1_{scn}$) and end ($\mathbf{l}^m_{scn}$).
2) \textit{Non-self-intersecting}:
In case the input stroke self-intersects, we pick its largest closed part starting and ending at the intersection.

\end{itemize}

\noindent
In this work, we regard lasso-selection as a mapping $f(P, V, L)$ that retrieves a subset of points $P_{s} \subseteq P$ based on the current viewpoint $V$ and lasso drawing $L$.
To be \emph{effective}, the mapping function $f$ should minimize difference between $P_{s}$ and target points $P_{t}$, \textit{i.e.},
%\vspace{-1mm}
\begin{equation}
	\underset{f}{\arg\min} \{d_J(P_{s}, P_{t}) \, | \, f(P, V, L) \rightarrow P_s\}
\end{equation}
\vspace{-2mm}

\noindent
% we employ Jaccard distance $d_J$ for measuring distance between $P_s$ and $P_t$ (see Sec.~\ref{ssec:eval}).
Meanwhile, we would also like the selection to be \textit{efficient}, 
% \emph{i.e.}, fast enough for fluid interaction, 
which requires the method should be fast enough for fluid interaction, 
and \textit{robust}, \emph{i.e.}, the performance remains effective and efficient over various conditions of point clouds ($P$), viewpoints ($V$), and lassos ($L$).

\begin{figure*}[htb]
\centering % avoid the use of \begin{center}...\end{center} and use \centering instead (more compact)
\includegraphics[width=0.95\textwidth]{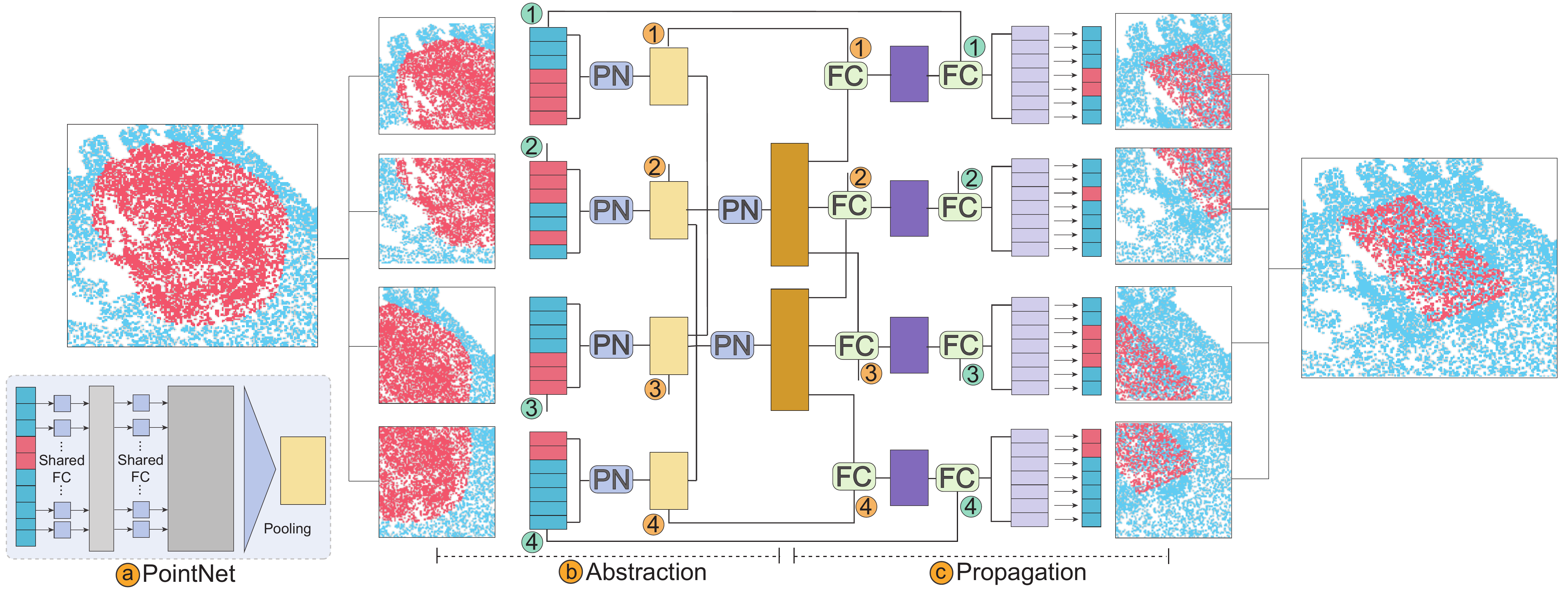}
	\vspace{-2.5mm}
\caption{Overview of network building. The DNN network is built upon (a) PointNet~\cite{Qi2017a}, and we employ a hierarchical structure that generates more local and global features using (b) abstraction and (c) propagation components. }
\vspace{-3mm}
\label{fig:archit}
\end{figure*}

\section{LassoNet}
\label{sec:lassonet}

% Given a point cloud $P$, a viewpoint $V$, and a lasso $L$, there exists a latent mapping $f(P, V, L)$ that models the user's intention of points selection. 
% Instead of employing dedicated heuristics to constrain the selection~\cite{Yu2012, shan_2012_interactive, Yu2016}, we opt to a deep neural network that can automatically learn an optimal mapping.
LassoNet pipeline (Fig.~\ref{fig:pipeline}) consists of three main stages:

\begin{enumerate}[leftmargin=*]

\item
\textbf{Interaction Encoding}.
A primary challenge for this work is implied by data heterogeneity, \emph{i.e.}, to associate viewpoint and lasso information with point cloud.
We address the challenge by \textit{3D coordinate transformation} that transforms point cloud from object space to camera space, and \textit{naive selection} that filters a subset of point cloud through CylinderSelection.
\vspace{-1mm}

\item
\textbf{Filtering and Sampling}.
The next challenge is to address scalability issue implied by great variability of point clouds (\emph{e.g.}, dense \emph{vs} sparse) and lasso selection (\emph{e.g.,} small \emph{vs} large).
We employ first an \textit{\cmo{intention} filtering} mechanism that filters points within an intention area, and then a \textit{farthest point sampling (FPS)} algorithm that \cmo{divides dense point clouds into partitions}.
\vspace{-1mm}

\item
\textbf{Network Building}.
Lastly, we build a hierarchical neural network to learn a latent mapping between point cloud, viewpoint, and lasso.
This network structure is inspired by PointNet++~\cite{Qi2017, Qi2017a} 
that achieves good performance for segmenting non-uniformly distributed \cmo{and varying density} point clouds.

\end{enumerate}

%%%%%%%%%%%%%%%%%%%%%%%%%%%%%%%%%%%%%%%%%%%%%%%
%%%%%%%%%%%%%%%%%%%%%%%%%%%%%%%%%%%%%%%%%%%%%%%
\subsection{Interaction Encoding}
\label{ssec:inter_encode}

% Our problem, unlike point cloud segmentation tasks, 
% is to infer a user's selection intention reflected by viewpoint and lasso.
To couple point clouds with viewpoint and lasso information is a prerequisite before we can train a DNN model.
We achieve this in two steps, as illustrated in Fig.~\ref{fig:pipeline}(b):

\begin{enumerate}[leftmargin=*]

% \vspace{1mm}
\item \emph{3D Coordinate Transformation}:
% Before lasso drawing on 2D surface, a user manipulates viewpoint to find an optimal angle of view for the target points.
The viewpoint is determined by many factors, including camera position and direction, FOV, projection type, etc.
% By saying optimal angle of view, the target points are viewable by the user, meanwhile not occluded by other points.
The information not available in point clouds, 
from which we only know point positions in 3D object space $\mathbb{R}^3_{obj}$.

We encode viewpoint information by transforming coordinates of all point clouds from $\mathbb{R}^3_{obj}$ to camera space $\mathbb{R}^3_{cam}$.
The transformation can be computed following the graphics pipeline.
Specifically, when a user draws the lasso, 
we record the current position and rotation of the camera, 
forming a 4$\times$4 projection matrix.
We then derive position of a point in the camera space $\textbf{p}_{cam}^i$ 
by multiplying the projection matrix with original position $\textbf{p}_{obj}^i$.
% $\textbf{p}_{cam}^i$ is still 3 dimensional vector.

% \vspace{1.5mm}
\item \emph{Naive Selection}:
% After finding an optimal viewpoint, the user can draw lassos on the 2D surface to enclose different regions of the point clouds.
The next question is how to associate a lasso with the point cloud. %, such that to infer the user's intention of selection region.
Notice that the lasso $L$ consists of an ordered list of points in 2D screen space $\mathbb{R}^2_{scn}$, while the point cloud has been transformed to 3D camera space $\mathbb{R}^3_{cam}$.

As indicated in Fig.~\ref{fig:pipeline}(b), we associate lasso information with the point cloud using a naive lasso selection method.
% The naive selection works in this way: 
First, we extrude a lasso $L$ in the 2D screen space to a frustum $F$ in 3D camera space, based on the current viewpoint and screen parameters.
Now both point cloud and lasso are transformed to camera space $\mathbb{R}^3_{cam}$.
Thus we can check if a point $\textbf{p}_{cam}^i$ is located inside $F$.
We append a binary value $w^i$ to indicate if $\textbf{p}_{cam}^i$ falls in $F$: 1 for inside (red points inside \autoref{fig:pipeline}(b)), and 0 for outside (blue points in \autoref{fig:pipeline}(b)).

\end{enumerate}

After these,
each point $\textbf{p}_{obj}^i$ is modeled as $\textbf{p}_{cam}^i := (x^i_{cam}, y^i_{cam}, z^i_{cam}, w^i)$, 
where $x^i_{cam}, y^i_{cam}, z^i_{cam}$ indicate position in camera space $\mathbb{R}^3_{cam}$, 
and $w^i$ indicates if the point falls inside the frustum $F$ extruded from a lasso.

% By this, we successfully associate point cloud with viewpoint and lasso information.
% This adaption not only enables effective learning of the latent mapping, but its simplicity also promotes real-time response that is crucial for interactive applications.

%%%%%%%%%%%%%%%%%%%%%%%%%%%%%%%%%%%%%%%%%%%%%%%
%%%%%%%%%%%%%%%%%%%%%%%%%%%%%%%%%%%%%%%%%%%%%%%
\subsection{Filtering and Sampling}
\label{ssec:filter_sample}

% To fulfill the \textit{robust} requirement, our method should be scalable for variability for both 
% 1) point cloud: total number of points in different datasets can range from few thousand (ShapeNet dataset) to hundreds of thousands (S3DIS and astronomy datasets);
% and 2) lasso selection: users may select a small part, e.g., an engine from an airplane in ShapeNet dataset, or a big part, e.g., a table in a room in experiment data S3DIS dataset.
To cope with varying scales implied by point cloud and lasso selection, 
one simple approach is to add more neurons of a neural network, \emph{i.e.}, scaling up the network width.
This, however, will greatly increase the computation time and wreck interactive response.
Instead, we employ a \textit{filtering}-and-\textit{sampling} approach as in Fig.~\ref{fig:pipeline}(c):

\begin{enumerate}[leftmargin=*]

% \vspace{1.5mm}
\item \emph{\cmo{Intention} Filtering}:
% Inspired by the concept of \textit{attention mechanism}~\cite{vaswani_2017_attention} 
% that has recently been popular in deep learning, 
We would like to filter out points which users definitely have no intention to select.
We employ a heuristic that \cmo{points distant from the lasso are less intended and can be filtered out.}
Ideally, the intention can be measured as a parabolic function based on distance to the lasso.

However, it can be inefficient to compute all point distances to a lasso.
Instead, we implement a method that is much simpler, yet gives results as good as the parabolic one.
When users draw a lasso on the 2D surface, 
we first find the lasso's bounding box (yellow dashed rectangle in Fig.~\ref{fig:pipeline}(c)).
We then expand the box 1.2 times outwards (solid purple rectangle in Fig.~\ref{fig:pipeline}(c)),
\cmo{since the points slightly outside the lasso could also be the intended targets.}
Points falling outside the expanded box are filtered out.
% Notice these computations are done in screen space, since measurements in 2D are much easier than those in 3D.

% \vspace{1.5mm}
\item \emph{Farthest Point Sampling (FPS)}:
Intention filtering can reduce the amount of points in a great extent.
However, in cases that the point cloud is dense or the lasso selection is big, 
there can still be too many filtered points \cmo{to fit into a single GPU memory}.
To cope with these situations, we further employ a downsampling algorithm of FPS 
\cmo{which can reduce the number of points meanwhile effectively preserve the convex hull of filtered points~\cite{eldar_1997_farthest}.
Here, we iteratively split the filtered points into multiple partitions, 
with each partition consisting of \cmo{up to $thre(FPS)$ points}.
All partitions are fed to the network for selection prediction.
The predictions of all partitions 
% from the network 
will be combined together as the final output.
% By this, we can fully leverage multiple GPUs to accelerate the computing.
}
% to reduce the number meanwhile effectively cover the entire set of filtered points~\cite{eldar_1997_farthest}. 

%%%%%%%%%%%%%%% CZT: I think the paragraph below can be removed %%%%%%%%%%%
% FPS first randomly selects a point $\textbf{p}^{s1}_{cam}$ from the filtered points, 
% then picks another point $\textbf{p}^{s2}_{cam}$ from the remaining points that is most distant to $\textbf{p}^{s1}_{cam}$,
% which yields a point set $\{\textbf{p}^{s1}_{cam}, \textbf{p}^{s2}_{cam}\}$.
% %, and the algorithm picks a third point $\textbf{p}^{s3}_{cam}$ that is most distant to $\{\textbf{p}^{s1}_{cam}, \textbf{p}^{s2}_{cam}\}$.
% This process is repeated iteratively until the number of sampled points reach the threshold.
% The distance here is measured as Euclidean distance in the camera space.

\end{enumerate}

By these,
a point cloud is divided into multiple partitions,
each of which consists of a set of sample points 
$\{\mathbf{p}^{si}_{cam} := (x^{si}_{cam}, y^{si}_{cam}, z^{si}_{cam}, w^i)\}_{i=1}^{thre(FPS)}$.
% where $thre(FPS)$ is set to 20,480 in this work.

\subsection{Network Building}
\label{ssec:net}

\textit{Filtering} and \textit{sampling} step ensures the amount of points is manageable by a neural network in real-time.
This, however, can greatly affect the prediction accuracy, which is then addressed by a hierarchical design of neural network as shown in~\autoref{fig:archit}.
  % to improve the accuracy.
A core component here is PointNet (PN)~\cite{Qi2017}, which directly consumes an unordered list of points and outputs a representative feature vector.
As illustrated in \autoref{fig:archit}(a), 
PN first employs a group of fully connected (FC) layers to map each point into a high-dimensional space.
The FC layers share parameter weights to reduce the number of parameters and accelerate network convergence.
% \cmo{
% \emph{Batch normalization} with a \emph{exponential weight decay}
% is applied to each FC layer to reduce overfitting.}
Finally, a pooling layer is used to aggregate the high-dimensional features 
and output a feature vector that can be regarded as the signature of the input point cloud.

% However, FC layers in PN still demand enormous parameters to model all points directly.
% Constrained by the capability of GPU devices, 
% PN applies sampling on the input points, 
% which is not desired since it decreases the prediction accuracy.
However, 
the pooling layer in PN only remains
the \emph{global} information of the whole point cloud.
% although PN can extract feature vector of the whole point cloud,
The relationship between a point and its \emph{local} neighborhood is missing,
which is not desired since it decreases the prediction accuracy.
Alternatively, we employ \textit{abstraction} and \textit{propagation} 
hierarchical structures as in PointNet++~\cite{Qi2017a} 
to generate features of both \emph{local} and \emph{global} information.

\begin{itemize}[leftmargin=*]

% \vspace{1mm}
\item
\textit{Abstraction} (\autoref{fig:archit}(b)):
We first divide the whole input points into several groups of equal size.
Each point group is represented as a level-0 feature vector.
We then apply PN to each point group, yielding a level-1 feature vector 
characterizing \textit{local} features for each group of points.
Each feature vector can actually be modeled as a set of high-dimensional points, 
which can again be processed by PN to extract correlations among point groups.
This grouping and correlation extraction processes are recursively repeated $k$ times.
By this, we obtain a level-$k$ feature vector that stores both global and local features of the input point cloud.

% \vspace{1mm}
\item
\textit{Propagation}:
The next challenge is how to propagate the level-$k$ feature vector to individual points.
% These features should be carefully propagated to avoid ineffective information.
% \cmo{For example, the information of the local context of the top-left conner of the point cloud
% should not be propagate to the points in the bottom-right conner.}
% To address this issue,
% we leverage the hierarchy built up \cmo{in the first stage} (\autoref{fig:archit}c).
% l在哪出现过？ bug
Here, we first concatenate level-$k$ with level-$(k-1)$ feature vectors, 
and applying a FC layer that generates a propagated layer of feature vector.
The process is again repeated $k$ times until each point group is propagated.
By this, 
we generated a final feature vector containing rich information for each point, 
including its local relation to neighboring points, 
and its global relation to the whole point cloud.
% \cmo{To avoid overfitting,
% we employ a \emph{dropout} layer with a drop rate of 0.3
% to the last FC layer.}
% To propagate a global feature\cmo{ $\textbf{f}_l$ }to a group from which it is extracted,
% we concatenate $\textbf{f}_l$ with the group's feature $\textbf{f}_{l-1}$,
% and then use a shared FC layers to generate a new feature (\autoref{fig:archit}d).
% The new feature are then used for further propagations.
% Again, this process are recursively repeated until the global features are propagated to each point.
% The final feature of each point contains rich information 
% from itself, to local neighborhood, to the whole point cloud.
% Now we can use this feature to determine whether the point is selected or not.
% We pass it to a group of shared fc layers to complete the \cmo{classification}.

\end{itemize}

Based on the final feature vector, 
we can predicate for each sample point $\mathbf{p}_{cam}^{si}$ 
a probability value $\rho^i$ using FC layers with a softmax function, which falls in [0, 1] indicating the probability that $\mathbf{p}_{cam}^{si}$ is selected.
% Finally, we replace $w$ for a sample point $\mathbf{p}_{cam}^{si}$ 
% with $\rho^i$.
If $\rho^i$ is larger than 0.5, we regard $\mathbf{p}_{cam}^{si}$ 
as selected; otherwise not.

\section{Model Experiments}
\label{sec:model}

% In order to solve the difficulties of traditional selection techniques in increasingly complex and diverse point cloud environments, 
% researchers start to use learning methods recently. 
% Most of the learning methods need a rich and comprehensive/exhaustive dataset to 
% train them and evaluate their effectiveness. 
To train LassoNet,
we collect a dataset of more than 30K lasso-selection records (Sec.~\ref{ssec:anno}) annotated on two publicly available point cloud corpora (Sec.~\ref{ssec:data_pre}).% with various scales and complexities. 
% We first describe the collection process and report statistics of the dataset.
Then we introduce the training process (Sec.~\ref{sec:training}), and finally report the quantitative evaluation results (Sec.~\ref{ssec:eval}).

%%%%%%%%%%%%%%%%%%%%%%%%%%%%%%%%%%%%%%%%%%%%%%%%%%%%%%%%%%%
%%%%%%%%%%%%%%%%%%%%%%%%%%%%%%%%%%%%%%%%%%%%%%%%%%%%%%%%%%%
\subsection{Point Cloud Preparation}
\label{ssec:data_pre}

We choose two point cloud corpora that have been widely used in many applications, \emph{e.g.}, robotics and scene reconstruction.
The first one is \emph{ShapeNet}~\cite{Yi2016ShapeNet}, 
containing in total over 16K point clouds of CAD models in 16 categories (e.g., airplane, car, bag).
Each point cloud consists of several thousand points \cmo{and point densities from 10K to 6M points/$m^3$}.
The points are divided into two to six parts, e.g., an airplane is divided into \emph{body}, \emph{wing}, \emph{engine}, and \emph{tail}.
The second corpus is Stanford Large-Scale 3D Indoor Spaces (S3DIS) dataset, 
containing 272 point clouds collected by high-resolution 3D scanning of rooms.
The point clouds exhibit a wide range of point numbers from 60K to 3M, \cmo{and point densities ranging from 0.2K to  60K points/$m^3$}.
The points are also divided into parts, e.g., chair, table, and floor. 

% \textbf{Pre-Processing.}
To improve the quality of annotation, we first filter out points in the following cases:
i) The whole point cloud consists of only one part, e.g., laptop and skateboard point clouds in ShapeNet;
ii) The points occluding the view heavily, e.g., ceiling points in S3DIS datasets.
Nevertheless, even after filtering, there are still too many point clouds in ShapeNet.
Thus, we further randomly select 15\% from each category.
% we conduct a pre-processing of the two point cloud corpora:
% \begin{itemize}
% 	\item For ShapeNet, we first remove two categories of point clouds (\emph{i.e.}, laptop and skateboard), since they only contains one or two parts of points, which are easy to be selected.
% 	Then, for each category, we randomly select up to 200 point clouds
% 	to balance the number of point clouds.
% 	\item For S3DIS, 
% 	we first remove the ceiling category to make the instances inside each point cloud easier to select. 
% 	Then, for the point clouds with more than 200K points,
% 	we reduce the number of points to 200K by random sampling to control the size of point clouds.
% \end{itemize} 
After filtering and sampling, we retrieve 2,332 point clouds in 14 categories from ShapeNet, and 272 point clouds from S3DIS.

\begin{figure}[t]
	\centering
	\includegraphics[width=0.95\columnwidth]{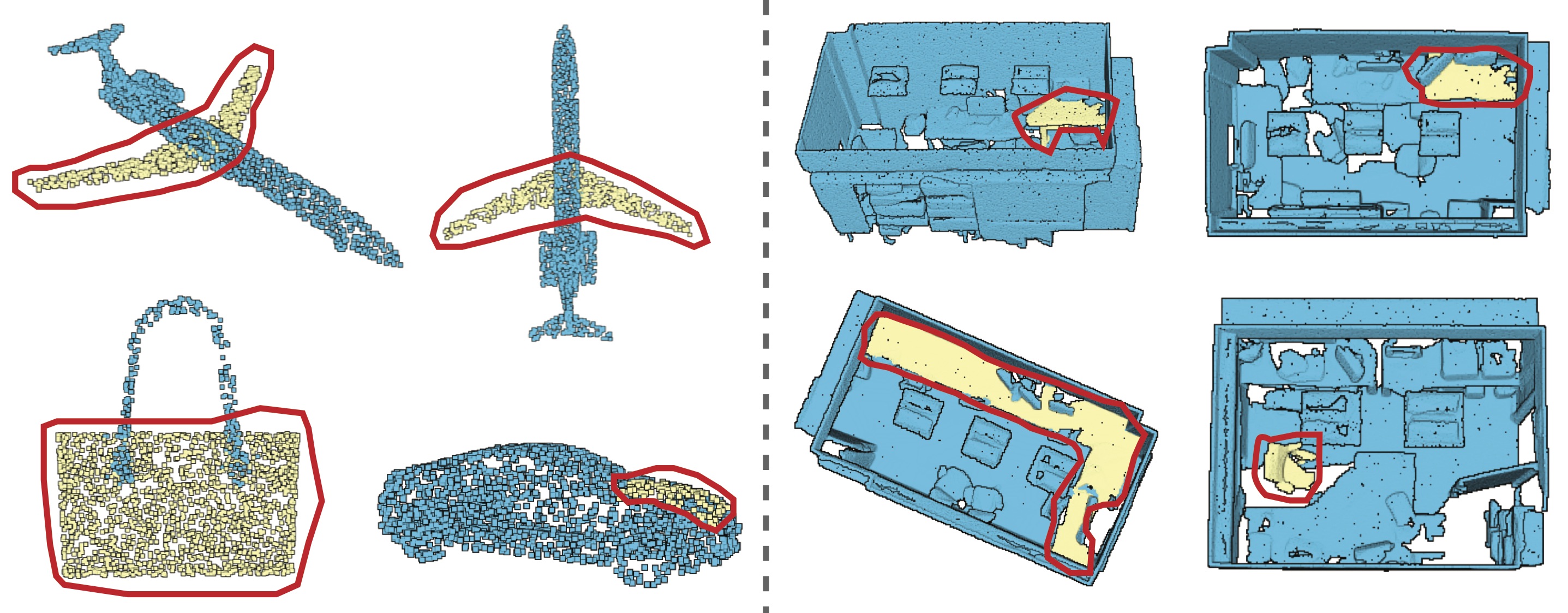}
	\vspace{-2mm}
	\caption{
		Exemplar annotation records for point clouds in ShapeNet (left) and S3DIS (right):
		target and interfering points are colored in yellow and blue respectively, while lassos are in red color.
		}
	\label{fig:pc_examples}
	\vspace{-2mm}
\end{figure}

%%%%%%%%%%%%%%%%%%%%%%%%%%%%%%%%%%%%%%%%%%%%%%%%%%%%%%%%%%%
%%%%%%%%%%%%%%%%%%%%%%%%%%%%%%%%%%%%%%%%%%%%%%%%%%%%%%%%%%%
\subsection{Lasso-Selection Annotation}
\label{ssec:anno}

We recruit 20 professional annotators to generate lasso selection records on the point cloud corpora.
The annotation is done on a web-based visualization platform that renders target points in yellow and interfering points in blue \cmo{with a fixed FOV of 60} (see Fig.~\ref{fig:pc_examples}).
For each point cloud, we randomly allocate one part (e.g., wings of an airplane, or a table in a room) as the target, and the others as interfering points.
The platform supports 5-DOF navigation using mouse input.

The annotators are asked to enclose target points by drawing an appropriate lasso (see red lassos in Fig.~\ref{fig:pc_examples}) from a good viewpoint.
Then, the target points inside of the lasso are highlighted to indicate how successful the selection is.
% Note that, we aim to find a latent mapping between viewpoint, input lasso and target points, so we do not mind to have interfering points inside of the lasso as long as the lasso can well enclose the targets.
Thus, for each set of target points, no matter the points are separated or not, we allow for only one lasso selection.
Taking the airplane in Fig.~\ref{fig:pc_examples} for an example, to select both wings, users are allowed to draw a lasso from different viewpoints as in the first two subfigures, but not to draw two lassos.
%\yu{Notice that each set of target points, no matter they are separated or not, only allows one lasso selection.}
We encourage the annotators to complete the selection as good as possible, so we do not set an explicit time limit in the annotation. 
% The annotators can either redraw a lasso if they feel not appropriate or continue to the next annotation task if a selection is fine.
When an annotation is finished, a backend process will record information of \textit{point cloud id}, \textit{target points ids}, \textit{current camera position \& direction}, and \textit{lasso drawings}.
To ensure annotation quality, we clean up records that cover less than 70\% of the target points or more than 80\% non-target points.

\begin{table}[h]
	\centering
	\caption{Statistics of lasso-selection records.}
	\label{table:selection_dataset}
	\begin{tabular}{lccc}
		\toprule
		 Dataset & $\#$Point Clouds & $\#$Targets & $\#$Records \\ \midrule
		 ShapeNet         & 2,332  &  6,297    & 19,432   \\ %\midrule
		 S3DIS         & 2,72   &  4,018    & 12,944       \\ 
		\bottomrule
	\end{tabular}
	\vspace{-2mm}
\end{table}

Table~\ref{table:selection_dataset} presents statistics of lasso-selection records.
In total, we have collected 19,432 lasso-selection records for 6,297 different parts of target points in ShapeNet point clouds, and 12,944 records for 4,018 different parts of target points in S3DIS point clouds.
Figure~\ref{fig:pc_examples} presents some examples of the annotations, which exhibit a wide range of diversities in:
1) \textit{point cloud} in terms of the whole (e.g., airplane, bag, rooms) and \textit{target points} (e.g., airplane wings, table, chair);
2) \textit{viewpoints} in terms of camera position (close by \emph{vs} far away) and angle (e.g., top, bottom, side);
3) \textit{lassos} in terms of position and shape.

%%%%%%%%%%%%%%%%%%%%%%%%%%%%%%%%%%%%%%%%%%%
\subsection{Network Training}
\label{sec:training}

%% TODO strong split
% We first introduce the training process with the loss function and important hyper parameters.
% Next we describe the evaluation metrics and report the results. 
%For evaluating the new method, and in particular the trained CNN, 
%we used k-fold cross-validation. In k-fold cross-validation, 
%the original sample is randomly partitioned into k equal sized sets. 
%In each of the k folds, a single set is retained as the validation data for testing the model, 
%and the remaining k$-$1 sets are used as training data. 
%In our evaluation, we set k=10.
% \subsubsection{Training}
% We separately train a model for each of $D1$ and $D2$,
% given the scales of them are far from the same.
% Nevertheless, LassoNet can easily adapt to them by adjusting 
% the FPS threshold and the hierarchical levels.

Following conventions in machine learning, we randomly split annotations records by point clouds into $9:1$ for training and testing.
In this way,
point clouds for testing do not appear in the training set.
This yields 2,092 out of 2,332 point clouds from ShapeNet, and 242 out of 272 from S3DIS for training.
% The remaining as testing data are used for evaluation.
% We conduct experiments on:
% 1) training two separate models for the datasets ($E_a$),
% and 2) training a single model for both two datasets ($E_b$).
% % Given the huge differences of point number in point clouds of ShapeNet (a few thousand) \emph{vs} S3DIS (hundreds of thousands), 
% % it is not feasible to train a single model using the same parameters for both annotations.
% % The same as PointNet++~\cite{Qi2017a}, we train two separate models for the annotations.
% We report the details of $E_a$ and discuss $E_b$ in Sec.~\ref{ssec:generalizability}.}
% \zw{this paragraph needs revision. Is $E_a$ referred later?}
% introduce detailed parameter settings in the models.

\vspace{1mm}
\noindent
\textit{Loss function.}
Since our task can be formulated as a per-point binary classification problem (\emph{i.e.}, selected \emph{vs} non-selected), we adopt a cross entropy loss function to train LassoNet.
For a training record, we calculate the loss on each point and then average the losses over all points to update the network by a backward propagation algorithm.
The loss for each training record can be calculated as:

\begin{equation}
	\mathcal{L} = - \frac{1}{n} \sum_{i=1}^{n} (\theta_0 s^i \log(\rho^i) + \theta_1 (1 - s^i) \log(1-\rho^i)),
\end{equation}

\noindent
where $n$ is the number of points in a training point cloud $P := \{\mathbf{p}^i\}_{i=1}^n$.
$s^i$ is a binary value indicating the ground-truth status of a point $\mathbf{p}^i$: 0 for interfering points, and 1 for target points.
$\rho^i$ is the probability value of $\mathbf{p}^i$ predicted by LassoNet.
To improve robustness of LassoNet on point clouds with extremely imbalanced numbers of target and interfering points, we add $\theta_0$ \& $\theta_1$ to control weights of the two classes.
Specifically, the interfering points are usually much more than target points in S3DIS annotations, thus we set $\theta_0 = 4$ and $\theta_1 = 1$.
In contrast, $\theta_0$ \&  $\theta_1$ are both set to 1 in ShapeNet.

\vspace{1mm}
\noindent
\textit{Hyper parameters.}
There are two hyper parameters that play important roles in LassoNet,
namely, threshold of FPS $thre(FPS)$, and
size of a group $size(g)$ 
% and number of groups $num(g)$.
in network building (Sec.~\ref{ssec:net}).

\begin{itemize}[leftmargin=*]
% \vspace{1mm}
\item
$thre(FPS)$ controls the maximum number of points fed into the network,
which depends on computational resource. % of GPUs. 
In our experiments, we use Nvidia 
% GeForce 
GTX1080Ti GPUs 
and set $thre(FPS)$ to 20,480.

% \vspace{1mm}
\item
$size(g)$ control the receptive fields of the network, ranging from 1 to $thre(FPS)$. 
A smaller $size(g)$ makes the network focus more on \textit{local} features of a point cloud, but leads to deeper hierarchy and more computational cost.
A bigger $size(g)$ allows the network to compute more efficiently, but less accurate predictions caused by the lack of sufficient local details.
$size(g)$should be set based on the characteristics of the target datasets.
Empirically, we set $size(g)$ to 2048 for ShapeNet annotations, since the point clouds contain only a few thousand points.
For S3DIS annotations, %which needs more local details to characterize instances,
we set $size(g)$ to 32 that strikes a good balance between effectiveness and efficiency.

% \item
% $num(g)$ determines the number of groups ranging from $\lceil \frac{thre(FPS)}{s_g} \rceil$ to $thre(FPS)$, which works in the same manner as $size(g)$, \emph{i.e.}, bigger will be more effectiveness but less computational efficiency.

\end{itemize}

\vspace{1mm}
\noindent
\textit{Implementation Details.}
% Initial tests confirmed that these hyper-parameters work well on the annotation datasets.
Adam optimizer is used to optimize the loss of the model.
We choose 0.9 for the momentum and 1e-3 for initial learning rate, 
which is reduced by half per 50 epoch.
To avoid overfitting, we employ batch normalization with a decay rate starting from 0.5 and exponentially grows to 0.99, and dropout with keep ratio of 0.7 on the last FC layer.
The models are implemented using TensorFlow 
and run on a server equipped with four NVIDIA 
% GeForce 
GTX1080Ti graphics cards. 
Each training process contains 200 epochs.

\subsection{Evaluation}
\label{ssec:eval}

\textbf{Accuracy performance}
is a main criterion for lasso selection techniques. 
As discussed in Sec.~\ref{ssec:prob}, 
the difference between selection points $P_s$ and target points $P_t$ should be minimized.
We measure the difference using Jaccard distance, which is calculated as:

\begin{equation}
d_J(P_s, P_t) = 1 - \frac{|P_s \cap P_t|}{|P_s \cup P_t|} = 1 - \frac{|P_s \cap P_t|}{|P_s| + |P_t| - |P_s \cap P_t|}
\end{equation}

We further include $F1$ score that is often used in measuring binary classification performance.
$F1$ is measured upon true positive (TP = $|P_s \cap P_t|$),
false positive (FP = $|P_s - P_s \cap P_t|$),
and false negative (FN = $|P_t - P_s \cap P_t|$): $F1 = 2TP / (2TP + FP + FN)$.
In general, $F1$ score tends to measure average performance, 
while $d_{J}$ tends to measure the worst case performance.
Both $d_{J}$ and $F1$ are in the range of [0, 1], 
where 0 indicates best performance for $d_{J}$ 
but worst performance for $F1$, and vice versa.

\begin{table}[!t]
	\centering
	\label{table:error_rate}
	\begin{tabular}{lcccccc}
		% \toprule
		\multirow{2}{0.9cm}{} & \multicolumn{3}{c}{ShapeNet} & \multicolumn{3}{c}{S3DIS} \\
		                        \cmidrule(l{2pt}r{2pt}){2-4} \cmidrule(l{2pt}r{2pt}){5-7}
		         & $d_J$ & F1 & \cmo{Time (ms)} & $d_J$ & F1 & \cmo{Time (ms)} \\ \midrule
		Cylinder    & 0.28 & 0.84 & \cmo{16.67} & 0.61 & 0.57 & \cmo{18.86}  \\
		LassoNet & 0.08 & 0.95 & \cmo{20.47} & 0.17 & 0.90 & \cmo{69.46} \\ 
		\bottomrule
	\end{tabular}
	\vspace{1mm}
	\caption{Performance of CylinderSelection and LassoNet on ShapeNet and S3DIS annotations.}
	\vspace{-6mm}
\end{table}

We compare LassoNet with CylinderSelection - a basic lasso-selection method for 3D point clouds.
Table 2 presents the comparison results on the testing annotations
from ShapeNet and S3DIS separately.
% Both of them only allow one lasso selection.
% Since LassoNet is based on the naive lasso selection method,
% we also report the performance of the naive lasso selection as benchmarks. 
% Compared with the naive lasso selection,
% LassoNet achieves significantly less error rate on both $D1$ and $D2$.
Overall, LassoNet achieves much better performance 
than CylinderSelection on both annotation datasets in terms of both $F1$ score and $d_J$.
Specifically, we notice that the performance of CylinderSelection drops much on S3DIS annotations, 
while LassoNet only drops a bit.
We hypothesis this is because S3DIS annotations are more diverse than ShapeNet annotations.
To validate the hypothesis, we conduct further evaluations from the following perspectives:

\begin{itemize}[leftmargin=*]

\begin{figure}[th]
	\centering
	\includegraphics[width=0.98\columnwidth]{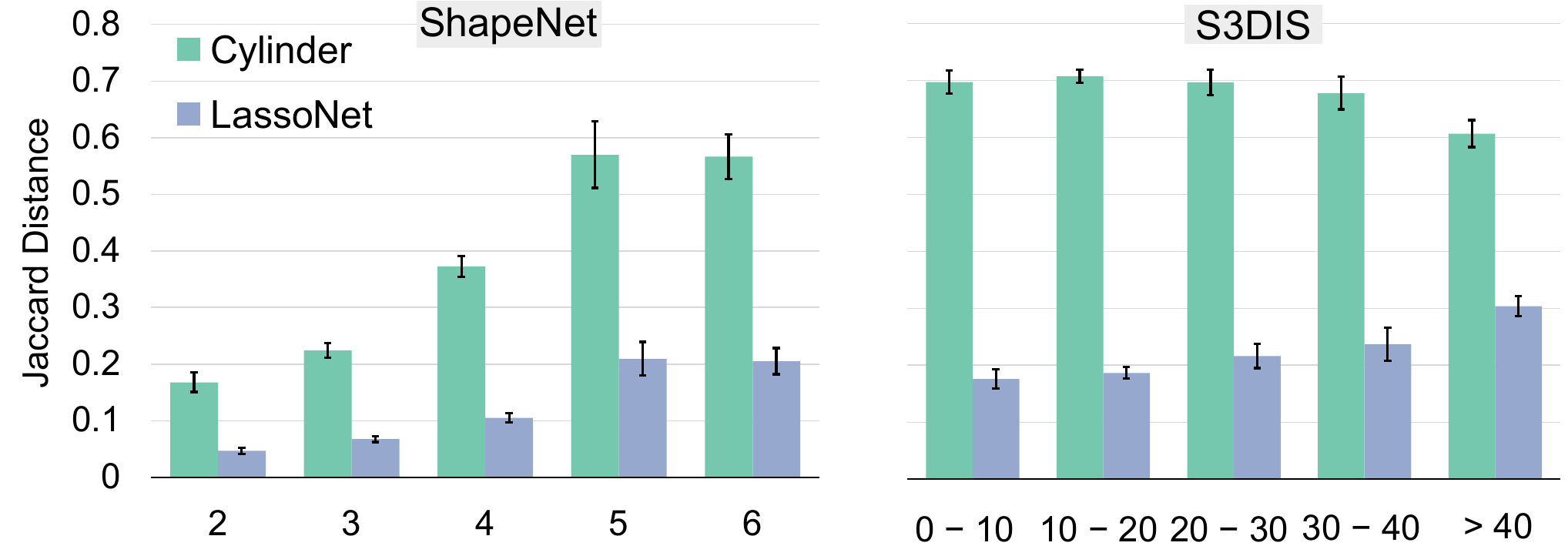}
	\vspace{-3mm}
	\caption{Jaccard distances of CylinderSelection and LassoNet measured upon scene complexity.}
	\label{fig:5-d1-results}
	\vspace{-4mm}
\end{figure}

% \vspace{1mm}
\item
\textit{Scene complexity}.
We quantify scene complexity using the number of parts in a point cloud.
Point clouds in ShapeNet contain a limited number of parts ($\leq$ 6), 
while S3DIS point clouds usually consist of tens of parts.
Figure~\ref{fig:5-d1-results} compares CylinderSelection 
and LassoNet over variations of scene complexity.
On the left, average $d_J$ are measured for ShapeNet annotations 
divided into groups of 2 $-$ 6 parts.
It can be observed that $d_J$ of CylinderSelection increases quickly to $\sim$0.58 when the number of parts increases to 5, while LassoNet remains to be less than 0.2. 
On the right, average $d_J$ are measured for S3DIS annotations divided into 
groups 
of [0, 10), [10, 20), [20, 30), [30, 40), and [40, +$\infty$) parts.
Again, $d_J$ for CylinderSelection remain high in all cases, 
while LassoNet remains low around 0.2.
Surprisingly, we notice that when the number of parts exceeds 40, 
$d_J$ of CylinderSelection drops, while LassoNet increases.

\begin{figure}[th]
	\centering	\vspace{-2mm}
	\includegraphics[width=0.985\columnwidth]{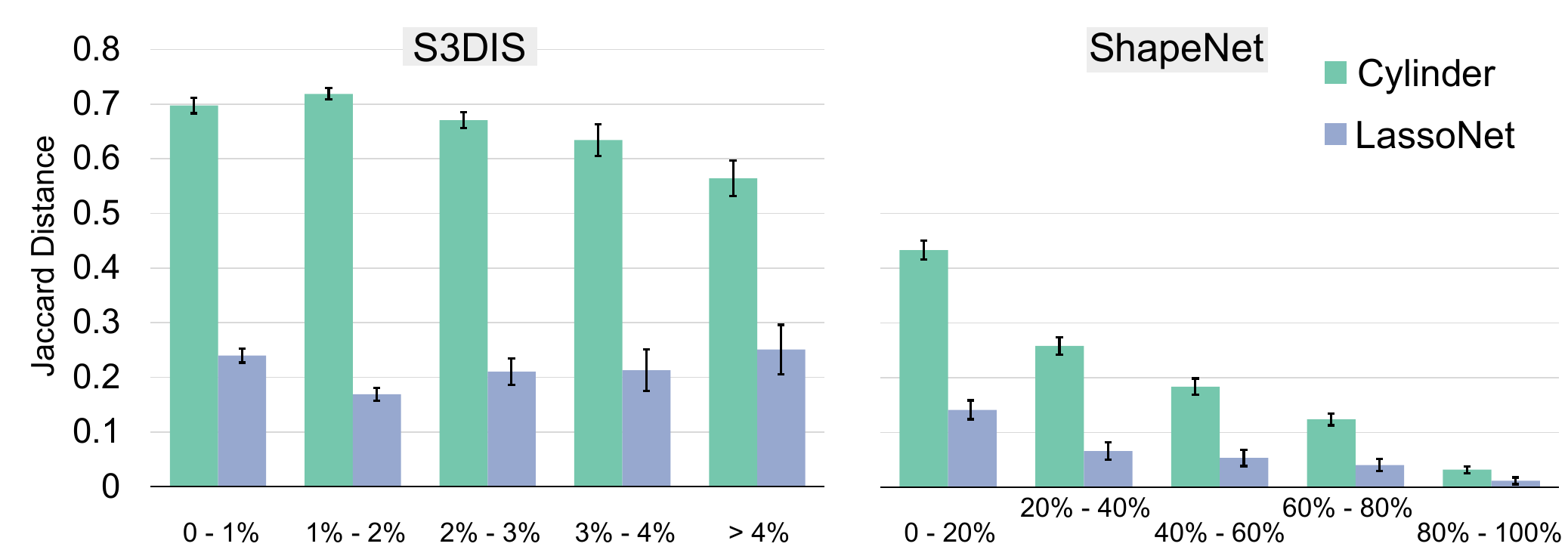}
	\vspace{-3mm}
	\caption{Jaccard distances of CylinderSelection and LassoNet measured upon task complexity.}
	\label{fig:5-d2-results}
	\vspace{-4mm}
\end{figure}

\vspace{1mm}
\item
\textit{Task complexity}.
We quantify task complexity using the percentage of target points in a point cloud.
Big targets (\emph{i.e.}, higher percentage) are typically easier to select than small ones (\emph{i.e.}, small percentage).
Figure~\ref{fig:5-d2-results} shows the comparison results.
Targets in ShapeNet are occupying higher percentages than those in S3DIS.
We divide ShapeNet annotations into 5 equal ranges, and measure the average $d_J$ as presented on the right.
As expected, $d_J$ of both CylinderSelection and LassoNet drop when the percentage of target points increase.
The same trend can be observed for CylinderSelection on S3DIS annotations, as shown Fig.~\ref{fig:5-d2-results}(left).
Since the scene is much complex, we divide the annotations according to target point percentages in the range of [0, 1\%), [1\%, 2\%), [2\%, 3\%), [3\%, 4\%), [4\%, +$\infty$).
In contrast, we can see that LassoNet achieves stable and better performances across the five groups.

\end{itemize}

% \noindent
% In summary, both scene and task complexity shield effects on the performance of CylinderSelection and LassoNet.
% Nevertheless, LassoNet achieves much better results in all scenarios.

\vspace{1mm}
\noindent
\textbf{Time performance} is another criterion for lasso selection techniques.
% We assess the computational efficiency of our system 
% by calculating the average time consuming to finish the prediction per selection.
Table 2 also presents a comparison of time costs for CylinderSelection and LassoNet on ShapeNet and S3DIS annotations.
Here, CylinderSelection is implemented in WebGL with average time costs of 16.67ms for ShapeNet and 18.86ms for S3DIS.
LassoNet requires additional times for network computation, which adds up to 20.47ms and 69.46ms for ShapeNet and S3DIS, respectively.
The increments are reasonable given that point clouds in S3DIS contains hundreds of times more points than those in ShapeNet.
The time costs are also comparable with state-of-the-art lasso-selection methods such as CloudLasso~\cite{Yu2012}.
Nevertheless, time costs for accomplishing accurate selection tasks of LassoNet are actually less than those of CylinderSelection and CAST; see Fig.~\ref{fig:result} and Sec.~\ref{sec:quan_results} for details.

\begin{figure}[th]
	\centering	
	% \vspace{-2mm}
	\includegraphics[width=0.95\columnwidth]{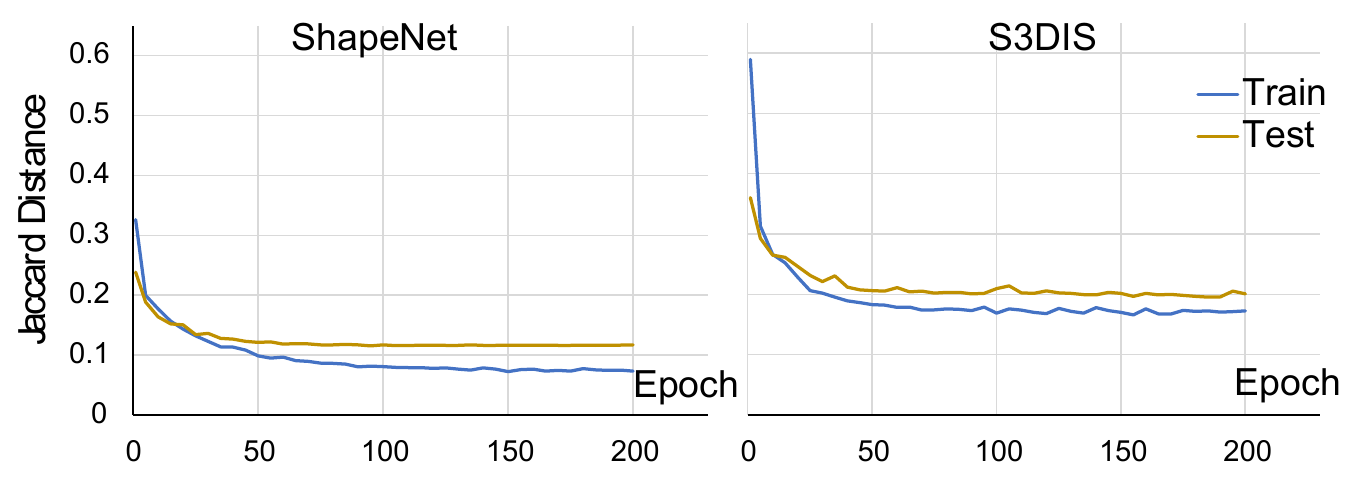}
	\vspace{-3mm}
	\caption{Jaccard distances per epoch in training and testing processes for ShapeNet (left) and S3DIS (right) annotations.}
	\label{fig:8-overfitting}
	% \vspace{-4mm}
\end{figure}

\vspace{1mm}
\noindent
\textbf{Underfitting and overfitting}
%Underfitting and overfitting 
are critical challenges in machine learning~\cite{Goodfellow-et-al-2016}.
Underfitting occurs when the model cannot fit the training set, while overfitting occurs when the model fits the training set
well but fails to fit the testing set.
In network training stage, we have adopted multiple strategies, including \emph{dropout}, \emph{weight decay}, and \emph{batch normalization}, to avoid the issues.
Nevertheless, to further investigate whether these issues occur, we examine $d_J$ per epoch in the training and testing processes, which are plotted as blue and red lines as shown in~\autoref{fig:8-overfitting}, respectively.
From the figures, we can notice that $d_J$ in training process decreases rapidly and smoothly, and $d_J$ in the testing process also decreases with a small gap between that in the training process.
The observations confirm that our model does not suffer from underfitting and overfitting problems.

\section{User Study}
\label{sec:study}

In reality, users typically complete selection of target points using a sequence of lassos.
To cope with this fact, we conduct a formal user study to further evaluate the performance of LassoNet in comparison with two lasso-selection methods of conventional \textit{CylinerSelection} and \textit{SpaceCast} $-$ a state-of-the-art density-based selection technique.
SpaceCast is chosen since it is the only method in the CAST family that is able to select a part of the cluster in case there is no density variation (such as two wings of an airplane).
Here, we allow users to refine a selection using Boolean operations of union, intersection, and subtraction for all three interactions.

This section reports quantitative results of the study in terms of \textit{efficiency} measured as completion time, and \textit{effectiveness} measured as Jaccard distance.
By comparing efficiency and effectiveness over different datasets, we further evaluate \textit{robustness} of LassoNet.

%%%%%%%%%%%%%%%%%%%%%%%%%%%%%%%%%%%
\subsection{Experiment Design}

\textbf{Participants:}
We recruited 16 participants (9 males and 7 females) in the study.
13 participants are students from different disciplines such as computer science and biochemistry, while the other three are research staff.
All participants had at least a Bachelor's degree.
The age of the participants ranges from 22 to 29, with the mean age of 24.69 years ($SD$ = 1.78).
All participants reported to be right-handed.
Four participants had experience of working with point clouds. 
Three participants had experience of manipulating 3D objects, and they are familiar with basic 3D interactions, including rotation and zoom-in/-out.
All participants completed the experiments in about 90 minutes.

\vspace{1mm}
\noindent
\textbf{Apparatus and Implementation:}
% \yu{It is not necessary to emphasis the fair comparison. Just say specifically what are the settings. If there is anything different, then you have to point them out and explain why we gave different settings.}
Testing datasets from ShapeNet and S3DIS were converted into a data format of point positions.
A web-based visualization is developed for LassoNet, while CylinderSelection and CAST are running on CAST application developed by the authors.
To eliminate bias caused by rendering effects, we adopted the same settings of FOV, background, and point colors with the CAST tool. 
Target points were rendered in orange color while interfering points and noise points were in blue.
LassoNet models ran on a backend server using one NVIDIA 
% GeForce 
GTX1080Ti graphics card.
All experiments were performed on a full HD resolution display (1920$\times$1080 px), with a standard mouse as the input device.

\begin{figure}[htb]
	\centering
	\includegraphics[width=0.98\columnwidth]{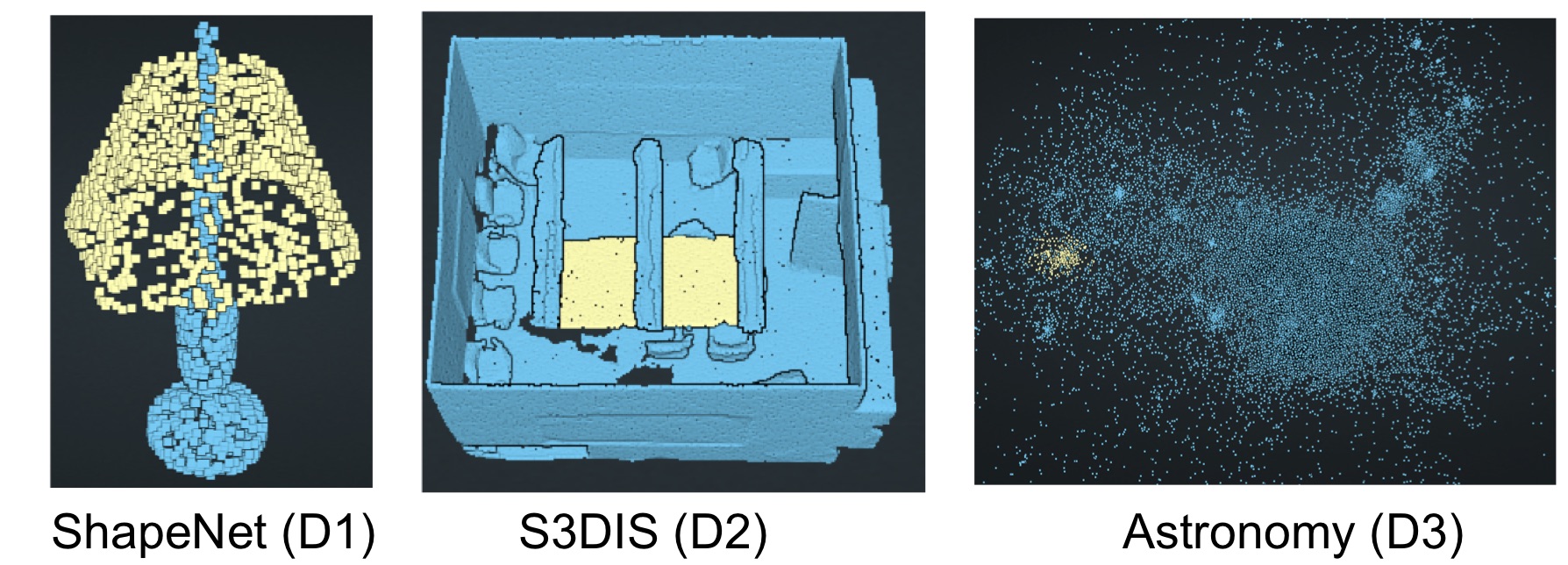}
	\vspace{-3mm}
	\caption{Exemplar point clouds employed in the user study.}
	\label{fig:userstudydata}
	\vspace{-4mm}
\end{figure}

\vspace{1mm}
\noindent
\textbf{Datasets:}
We conducted the user study on three different datasets, as shown in Fig.~\ref{fig:userstudydata}.
Besides ShapeNet (denoted as \textit{D1}) and S3DIS (\textit{D2}) described in Sec.~\ref{ssec:data_pre}, we recruited a third dataset of astronomy point clouds that were used in CAST experiments (\textit{D3}). 
D3 consists of four point clouds, each of which is made up of 200K to 400K points representing multiple particle clusters.
The clusters have equal uniform densities and are surrounded in a low-density noise environment. The target cluster was located either in the center or was partially surrounded by interfering points so that it is tricky to find a clear view to the whole target.
% One point cloud in D3 is a cosmological N-Body simulation dataset with a high-density core in the center and many small and high density clusters around.
We trained a new model for D3, using only 600 lasso-selection records manually annotated by ourselves.
The other settings are the same as those used when training S3DIS annotations.
Same as~\cite{Yu2016}, participants were asked to select some of the small clusters.
From each dataset, we selected three different point clouds with one meaningful part as target points.
All the point clouds and target points were not used for training. 
In total, there were 9 assignments (3 point clouds $\times$ 3 targets) for participants to complete using each method.

% These datasets were designed to have different features to more comprehensively evaluate: 

% d1: The d1 comes from ShapeNet Part Dataset. Number of points is from 2000 to 3000. Each scene is a single object. The target clusters are wings in an airplane(seperate target), a body in car(refining contour) and canopy in lamp(internal cross with interfering). 

% d2: The d2 comes from S3DIS. Number of points is 20w. Each scene is a room comprised with chairs, tabels, wall, etc. The target are table(big,forward occlusion), chair(small) and bookcase(big and close connect with interfering). 

% d3: The d1 comes from CAST~\cite{Yu2016}. Number of points if from 20w to 40w. Clusters and Shell are synthetic,they have same uniform density. Simulation is a cosmological data from real world. The target...

\vspace{1mm}
\noindent
\textbf{Task and Procedure:}
The task was to select target points marked in orange while avoiding interfering points marked in blue.
Selected points would be marked in red. 
In CylinderSelection and SpaceCast, the participants were allowed to refine the selections by three Boolean operations: union, intersection, and subtraction, in case they were not satisfied by the results.
% Moreover, in the LassoNet, the participants were allowed to zoom-in/-out the camera.
The participants were reminded that completion time is also an evaluation metric. 
So they were expected to complete the tasks as soon as possible in case they were satisfied with the results.
% We also reminded them do not aim at a perfect selection using a lot of refinements.
The participants could take a 5-minute break when they felt tired.

Before actual experiments, we explained to the participants about the principles of the next lasso-selection method.
We demonstrated how to change the viewpoint, draw lassos, and select the target points on the screen.
To ensure the participants fully understood the interactions, they were asked to practice with three training datasets. 
In the training trials, we gave the participants as much time as they needed.
To suppress learning effects gained from previous assignments, we assigned a sequence of lasso-selection methods pseudo-randomly to each participant.
When participants felt satisfied with the results, they proceeded to the next task by pressing a \textit{Submission} button, and a backend process automatically recorded the completion time and accuracy for the current task.
In the end, the participants were asked to complete a questionnaire for user feedback on their satisfactory of each method.
%The questions include if they feel the lasso-selection methods are difficult to learn and to use, if the selected points match with their intension, and advices for improvements.

\begin{figure}[b]
\vspace{-4mm}
	\centering
	\includegraphics[width=0.98\columnwidth]{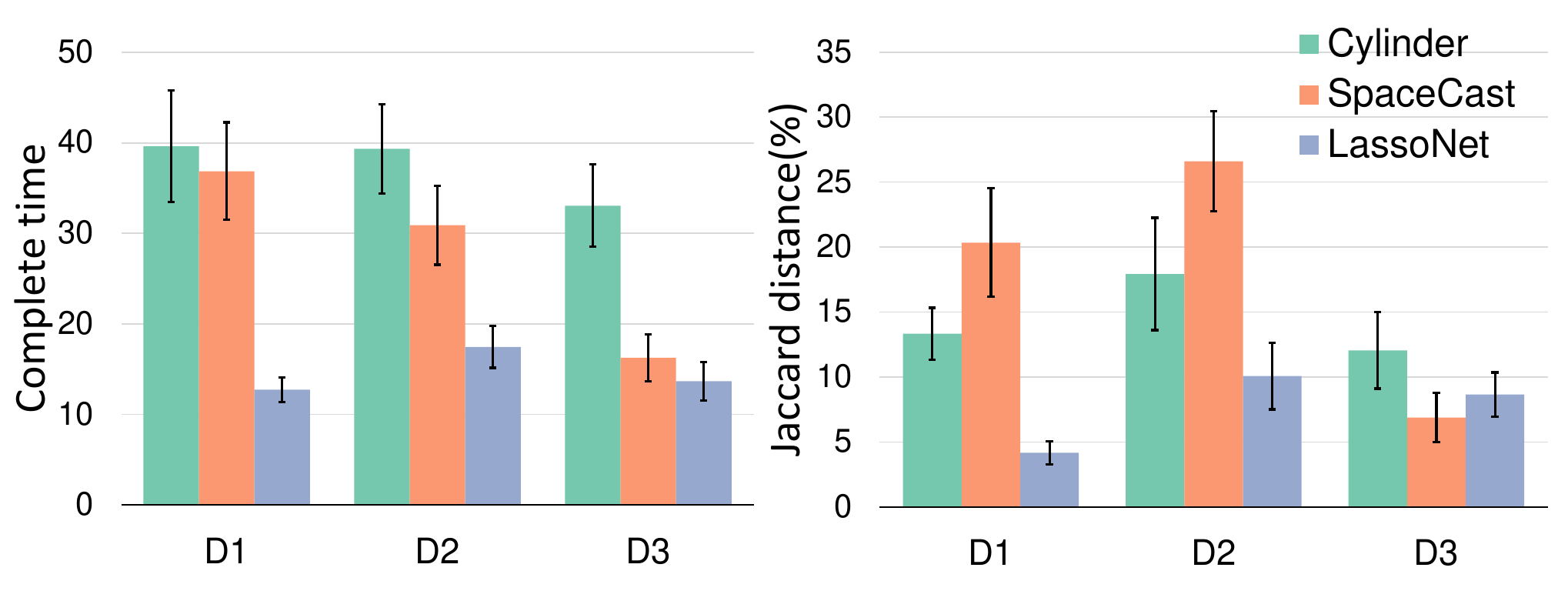}
	\vspace{-4mm}
	\caption{Comparison of completion time (left) and Jaccard distance (right) of CylinderSelection, SpaceCast, and LassoNet on three datasets.}
	\label{fig:result}
	% \vspace{-4mm}
\end{figure}

%%%%%%%%%%%%%%%%%%%%%%%%%%%%%%%%%%%%%%%%%%%%%%%
%%%%%%%%%%%%%%%%%%%%%%%%%%%%%%%%%%%%%%%%%%%%%%%

\begin{figure*}[htb]
	% \vspace{-4mm}
	\centering 
	\includegraphics[width=1.9\columnwidth]{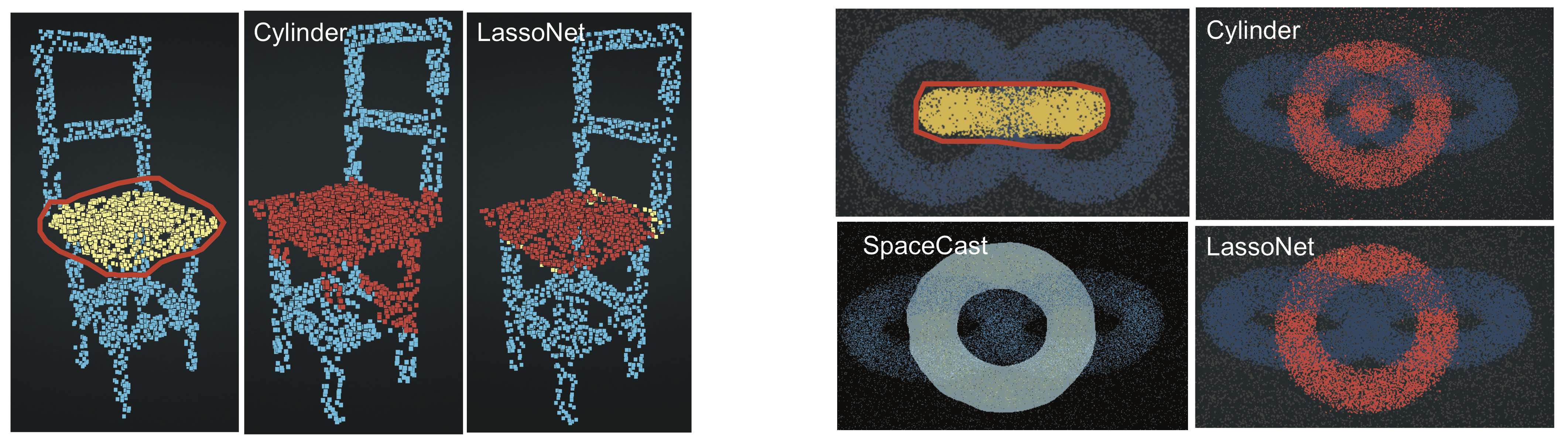}
	\vspace{-3mm}
	\caption{
	Comparison of LassoNet with prior methods on two different examples. (Left) LassoNet \textit{vs} CylinderSelection on a chair in ShapeNet dataset. (Right) LassoNet \textit{vs} CylinderSelection \textit{vs} SpaceCast on an artificial dataset.}
	\vspace{-4mm}
	\label{fig:results_example1}
\end{figure*}

\subsection{Hypotheses}
\label{ssec:hypo}

We expected LassoNet would outperform CylinderSelection on all three datasets in terms of completion time and Jaccard distance.
We also expected LassoNet would achieve similar performance with SpaceCast on D3, while better performances on D1 and D2 which do not have varying point density.
Moreover, since CylinderSelection does not require any additional computation, we expected that CylinderSelection would achieve similar performance on all three datasets.%, whilst heuristic-based SpaceCast would be the least generalizable method.

% We expected that LassoNet would outperform Naive and SpaceCast methods in all three evaluation criteria:
% \yu{the hypotheses, will it be a good idea to have something like: 1. for D1 and D2, (describe the datasets), LassoNet would be more efficient and effective, i.e.  However CAST and CylinderSelection would not have too much difference. 2. for D3, which has density differences, SpaceCast would be more efficient and effective, however LassoNet would not be so different.}

\begin{itemize}[leftmargin=*]
	
	% \vspace{0.5mm}
	\item
	\textit{H1}: LassoNet would be more \textit{efficient}, \emph{i.e.}, less completion time, than CylinderSelection on all datasets (H1.1). 
	Compared with SpaceCast, LassoNet would be more efficient on D1 \& D2, while equally efficient on D3 (H1.2).
	
	% \vspace{0.5mm}
	\item
	\textit{H2}: 
	LassoNet would be more \textit{effective}, \emph{i.e.}, smaller Jaccard distance, than CylinderSelection on all datasets (H2.1). 
	Compared with SpaceCast, LassoNet would be more effective on D1 \& D2, while being equally efficient on D3 (H2.2).
	
	% \vspace{0.5mm}
	\item
	\textit{H3}:
	%一个问题，一致得差能叫鲁棒性好吗。yzg
	CylinderSelection would be the most \textit{robust}, \emph{i.e.}, similar completion times (H3.1) and Jaccard distances (H3.2), on all three datasets.
	LassoNet would be more robust than SpaceCast. %\yu{re-consider about generalizable}
	
\end{itemize}

%%%%%%%%%%%%%%%%%%%%%%%%%%%%%%%%%%%%%%%%%%%%%%%
%%%%%%%%%%%%%%%%%%%%%%%%%%%%%%%%%%%%%%%%%%%%%%%
\subsection{Quantitative Results}
\label{sec:quan_results}

We collected in total 432 records (16 participants $\times$ 3 techniques $\times$ 9 assignments) from the user study.
Figure~\ref{fig:result} presents a comparison of mean completion time (left), Jaccard distances (right), and their 95\% confidence intervals for each technique conducted on each dataset.
At noticed, LassoNet outperformed CylinderSelection on both completion time and accuracy.
LassoNet also achieves better performance than SpaceCast on D1 \& D2, and similar performance on D3.

We performed a two-way ANOVA (3 techniques $\times$ 3 datasets) on both completion time and Jaccard distance.
Before the analysis, we first confirmed that all results of completion time and Jaccard distance in each condition follow normal distribution using a Shapiro-Wilk test.
Prerequisites for computing ANOVA are fulfilled for the hypothesis.
All hypotheses are confirmed by the analyses.
Below we report details of individual analysis.

% \begin{table}[htb]
% 	\caption{Mean IoU, F1, and completion time per technique and dataset. }
% 	\centering
% \addtolength{\tabcolsep}{-3pt}
% 	\begin{tabular}{c|cc cc|cc} \toprule
% 		Technique & IoU & 0.95 CI & F1 & 0.95 CI & Time & 0.95 CI \\ \midrule
% 		\multicolumn{1}{c}{} & \multicolumn{6}{c}{Part Dataset}     \\ \midrule
% 		Cylinder  & 86.6 &[84.5, 88.8] & 92.6 &[91.3, 93.9] & 39.6 &[30.6, 48.6]  \\
% 		SpaceCast & 79.6 &[75.1, 84.1] & 87.9 &[84.9, 90.8] & 36.8 &[30.9, 42.8]  \\
% 		Ours      & 95.8 &[95.2, 96.4] & 97.9 &[97.5, 98.1] & 12.7 &[10.6, 14.8]  \\ \midrule
% 		\multicolumn{1}{c}{} & \multicolumn{6}{c}{S3DIS Dataset}    \\ \midrule
% 		Cylinder  & 82.0 &[75.7, 88.5] & 88.9 &[84.2, 93.7] & 39.3 &[31.5, 47.1] \\
% 		SpaceCast & 73.4 &[69.7, 77.1] & 83.5 &[80.7, 86.4] & 30.9 &[24.3, 37.5] \\
% 		Ours      & 89.9 &[87.4, 92.4] & 94.3 &[92.7, 95.9] & 17.4 &[14.1, 20.8] \\ \midrule
% 		\multicolumn{1}{c}{} & \multicolumn{6}{c}{Cast Dataset}     \\ \midrule
% 		Cylinder  & 87.9 &[83.7, 92.2] & 92.9 & [89.9, 95.8] & 33.0 &[26.5, 39.6] \\
% 		SpaceCast & 93.1 &[90.9, 95.2] & 96.1 & [94.4, 97.8] & 16.2 &[12.4, 20.0] \\
% 		Ours      & 91.3 &[89.1, 93.5] & 95.3 & [93.9, 96.7] & 13.6 &[10.8, 16.4] \\ \midrule
% 	\end{tabular}
% \addtolength{\tabcolsep}{3pt}
% \end{table}

% 60 results are generated for each experiment condition (\textit{selection techniques} $\times$ \textit{datasets}).\zw{need to clarify this earlier}

%%%%%%%%%%%%%%%%%%%%%%%%%%%%%%%
%%%%%%%%%%%%%%%%%%%%%%%%%%%%%%%
\vspace{1mm}
\noindent
\textbf{Completion Time}.
% \label{ssec:comp_time}
As expected, \textit{selection technique} had a significant effect on completion times ($F_{2,429}$ = 82.73, p \textless .0001). 
Average completion times (Fig.~\ref{fig:result}(left)) are 37.34s for CylinderSelection (SD = 18.74), 27.14s for SpaceCast (SD = 17.72), and 15.49s for LassoNet (SD = 7.81). 
Post-hoc tests using Bonferroni correction indicate that LassoNet is significantly faster than both CylinderSelection (p \textless .0001) and SpaceCast (p \textless .0001).
Through more detailed probes, we found that LassoNet was significantly faster than CylinderSelection ($F_{1,47}$ = 69.62, p \textless .0001), ($F_{1,47}$ = 61,51, p \textless .0001), ($F_{1,47}$ = 39.60, p \textless .0001) on D1, D2, and D3, respectively.
This confirms the hypothesis H1.1.
% Moreover, 
We also found that LassoNet was significantly faster than SpaceCast ($F_{1,47}$ =72.28 , p \textless .0001) on D1, and ($F_{1,47}$ = 28.28 , p \textless .0001) on D2. 
No significant difference is observed for LassoNet and SpaceCast on D3 ($F_{1,47}$ =2.30 , p = 0.13). 
This further confirms the hypothesis H1.2.
% When we analyzed the results for each dataset separately, Ours is significant faster than Cylinder(F=69.62, p\textless .0001) and SpaceCast(F=72.28, p\textless .0001) in d1. 
% Similarly in d2, Ours is significant faster than Cylinder(F=61.51, p\textless .0001) and SpaceCast(F=28.28, p\textless .0001). 
% While we did not observe a significant improvement of ours in d3 (F = 2.30, p = 0.13) compared with SpaceCast, ours slightly outperforms it about 2.59s on average. 

We checked effects of \textit{dataset} on completion time.
Using Bonferroni correction test, we found that \textit{dataset} shields a significant effect on SpaceCast ($F_{2,141}$ = 31.72, p \textless .0001), while no significant effect on ClyinderSelection ($F_{2,141}$ = 1.91, p = 0.15) and LassoNet ($F_{2,94}$ = 1.91, p = 0.008).
This result confirms the hypothesis H3.1.

% Furthermore, the SD of three techniques confirmed H3.
% We found a significant effect of dataset on completion time (F = 11.98, p \textless .0001). 
% The participants required more time to select in d1 and d2. Average completion times are 29.65 s for d1 (SD = 19.30), 29.23 s for d2 (SD = 16.83), 20.99 s for d3 (SD = 14.33). 

% When we analyzed the results for each dataset separately, Ours is significant faster than Cylinder(F=69.62, p\textless .0001) and SpaceCast(F=72.28, p\textless .0001) in d1. 
% Similarly in d2, Ours is significant faster than Cylinder(F=61.51, p\textless .0001) and SpaceCast(F=28.28, p\textless .0001). 
% While we did not observe a significant improvement of ours in d3 (F = 2.30, p = 0.13) compared with SpaceCast, ours slightly outperforms it about 2.59s on average.
%%%%%%%%%%%%%%%%%%%%%%%%%%%%%%%
%%%%%%%%%%%%%%%%%%%%%%%%%%%%%%%
\vspace{1mm}
\noindent
\textbf{Jaccard Distance}.
We repeated ANOVA tests on Jaccard distance (Fig.~\ref{fig:result}(right)), by which significant effects imposed by selection techniques were observed ($F_{2,429}$ = 29.77, p \textless .0001).
LassoNet achieved a much lower mean Jaccard distance of 7.65\% (SD = 6.94), in comparison with CylinderSelection (mean = 14.44\%, SD = 11.61) and SpaceCast (mean = 17.95\%, SD = 14.68).
LassoNet is significantly more effective than CylinderSelection on all three datasets (F = 36.38, p \textless .0001), which confirms the hypothesis H2.1.
Compared to SpaceCast, LassoNet is slightly more but not significant (F=1.87, p=0.18) error-prone on D3, whilst it is significantly effective on D1 (F = 72.28, p \textless .0001) and D2 (F = 28.28, p \textless .0001).
These results confirm H2.2.

Though not significant, CylinderSelection achieves the most consistent Jaccard distances on different datasets ($F_{2,141}$ = 3.52, p \textless 0.05), in comparison to SpaceCast ($F_{2,141}$ = 32.60, p \textless .0001) and LassoNet ($F_{2,141}$ = 10.73, p \textless .0001).
Nevertheless, LassoNet is more stable than SpaceCast.
Hypothesis H3.2 hereof is confirmed.

% \textit{Datasets} also shield significantly effect on error rates (F = 21.64, p \textless .0001), with ($F_{2,30}$ = 3.52, p \textless 0.05) for naive, ($F_{2,30}$ = 10.73, p \textless .0001) for LassoNet , and ($F_{2,30}$ = 32.60, p \textless .0001) for SpaceCast.  
% Because of the complexity of scene, the mean error rates in d1(M=12.63) or d2(M=18.21) are higher than if in d3(M=9.21). 

%When we analyzed the results for each dataset separately, Ours is significant more accurate than Cylinder(F=69.62, p\textless .0001) and SpaceCast(F=72.28, p \textless .0001) in d1. 
%Similarly in d2, Ours is significant more accurate than  Cylinder(F=61.51, p \textless .0001) and SpaceCast(F=28.28, p \textless .0001). 
%While data showed that in d3 Ours was slightly more error prone than SpaceCast, this difference was not statistically significant(F=1.87, p=0.18). 
%Both F1 of them are more than 95\%, therefore, close to a perfect result. 

%%%%%%%%%%%%%%%%%%%%%%%%%%%%%%%
%%%%%%%%%%%%%%%%%%%%%%%%%%%%%%%
\subsection{Qualitative User Feedback}

We also collected qualitative user feedback from the participants after the user study. 
\cmo{13 out of 16 participants prefer LassoNet, due to its simplicity to learn and to use.}
One participant stated that ``a person knowing how to control mouse should feel no difficulty in lasso-selection".
They also appreciated that the visual interface returned immediate feedback upon lass selections.
The Boolean operations of union, intersection, and subtraction posed some difficulty for them in the beginning, but they fully understood the operations through several trial-and-error trainings finally.
Below we summarize their feedback for each method.

\begin{itemize}[leftmargin=*]
	
	% \vspace{1mm}
	\item
	\textit{CylinderSelection}.
	All participants felt that results from CylinderSelection are most predictable.
	Some participants reported that this was highly appreciated because by then they can refine the selections using Boolean operations as expectations.
	However, we also noticed that the participants showed interests to refine selections at the beginning, but the interests dropped quickly after a few assignments.
	This reaction was particularly obvious on D2, where the scenes are complex so that participants would need to change viewpoints very often when making refinements.
	This explains why average completion time of CylinderSelection is slightly less on D2 than on D1. 
	
	% \vspace{1mm}
	\item
	\textit{SpaceCast}.
	Three participants expressed high praise for SpaceCast on D3, which allowed them to make pretty accurate selections.
	In fact, most participants would choose SpaceCast as their favorite method if the experiments were conducted on D3. 
	However, all participants felt that SpaceCast was very unpredictable on D1 \& D2, even though we had clearly explained the underlying mechanism.
	Often the results were only a part of what they intended to select.
	For instance, when the assignment was to select the left wing of an airplane (see Fig.~\ref{fig:lassoselections}), SpaceCast often selected only the engine part. We suspect the reason was that the engine has a slightly higher density of points than the wing. 
	
	% \vspace{1mm}
	\item
	\textit{LassoNet}.
	Most participants favored LassoNet since the selections best match with their intention.
	``It seems the method can really understand what I want", one participant commented.
	Nevertheless, the participants also figured out that refinement using LassoNet is not as feasible as CylinderSeleciton.
	When making refinements, participants often select only a few points at a very close view.
	In such scenarios, LassoNet tends to select more points that exhibit strong correlations with the target points (e.g., neighboring, symmetric, etc.), see Fig.~\ref{fig:good_bad} for an example.
	They suggested adding Boolean operations in the technique for refinements.
	
\end{itemize}

\section{Discussion}
\label{ssec:discussion}

%%%%%%%%%%%%%%%%%%%%%%%%%%%%%%%
\subsection{Examples}

Figure~\ref{fig:teaser} presents a typical example of selection task: in a complex room, users need to three regions of different objects.
For conventional CylinderSelection method, users need to adjust viewpoints according to the current region of selection, and also need to refine selections using Boolean operations.
Instead, LassoNet can complete the task directly from the top view.
Even though the targets are partially occluded, LassoNet can still correctly deduce regions of user intention based on the viewpoint and lassos.

Figure~\ref{fig:results_example1} presents the comparison of LassoNet with prior methods on two different point clouds.
The left side presents a chair in ShapeNet dataset.
The task is to select the seat of the chair.
The viewpoint is changed to a good position that allows users to draw a lasso.
Obviously, from this view direction, CylinderSelection selects the seat and parts the legs, which is not desired.
In comparison, LassoNet successfully separates the seat from legs and produces a good selection result.
On the right side, the task is to select one from three interlocking rings by drawing a lasso from the viewpoint, as shown in the top-left corner.
As expected, CylinderSelection selects all points inside of the lasso, whilst LassoNet and SpaceCast achieve equally good results.

\begin{figure}[t]
	\centering 
	\includegraphics[width=0.9\columnwidth]{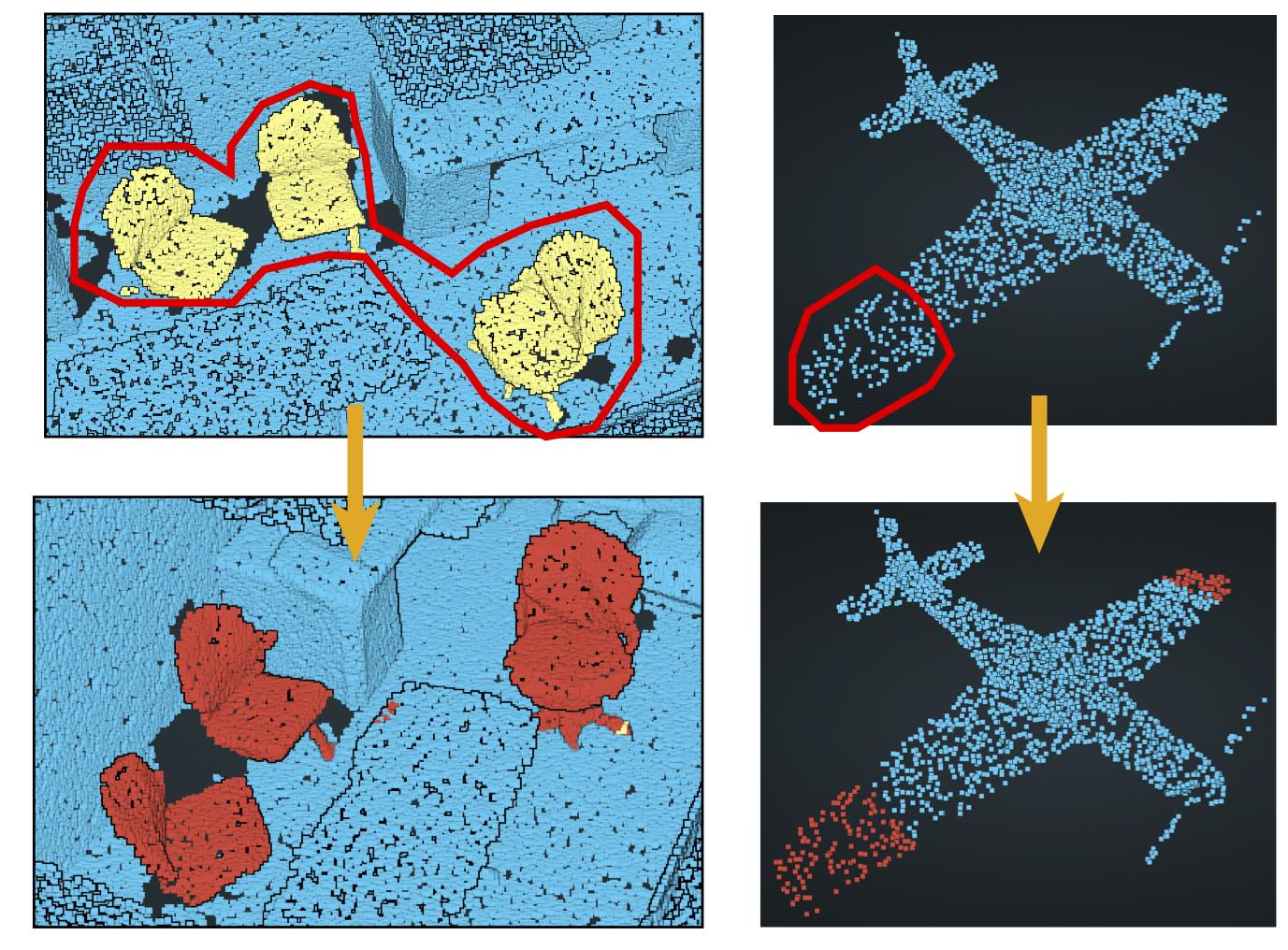}
	\vspace{-2mm}
	\caption{
	Typical examples of a good (left) case and a bad (right) case.
	}
	\vspace{-2mm}
	\label{fig:good_bad}
\end{figure}

LassoNet can also effectively select multiple separate parts by using only one lasso, as illustrated in Fig.~\ref{fig:good_bad} (left).
Here, a user draws a lasso to enclose all three target chairs, and LassoNet correctly segments points belonging to the chairs from surrounding points.
Notice that the training dataset does not include such cases (for S3DIS data, only one object is selected as the target for each annotation), LassoNet (probably) identifies similarity features among the chairs and select all of them together.
However, we also notice that sometimes LassoNet gives unexpected results, as shown in Fig.~\ref{fig:good_bad} (right).
We assume that LassoNet (again probably) detects symmetric features of trailing edge in the left and right airfoils, which is however not desired.

% The interaction of LassoNet is as simple as other lasso selection techniques: all of them request an input lasso and select points based on their unique heuristic. 
% The advantage of LassoNet is that, it does not rely on the density distribution of the data. 
% Thus, in case users give a lasso which is sufficient cover the target cluster, which does not have an intrinsic meaning of density, LassoNet is still able to have the whole cluster selected. 
%An even more interesting feature of LassoNet is that it is flexible to select half of the target cluster:
%in case the input lasso only covers half of a cluster, however, comparing to other clusters the lasso clearly indicate the intended cluster, then half of the intended cluster would be selected. 
%On this point, the CAST family provides three selection techniques: SpaceCast is able to select part of a cluster, however, in case users intend to select the whole cluster, then they need to draw a slight bigger lasso than the target cluster since SpaceCast only considers the points inside of the lasso; both TraceCast and PointCast treat clusters as individual objects, so they are unable to select part of a cluster without helping from Boolean Operations. 
% Thus, the robustness of LassoNet brings in an additional benefit that is its simplicity to learn and to use.

%%%%%%%%%%%%%%%%%%%%%%%%%%%%%%%
\subsection{Generalizability}
\label{ssec:generalizability}

% \begin{table}[!h]
% 	\centering
% 	\label{table:error_rate2}
% 	\vspace{-2mm}
% 	\begin{tabular}{lcccc}
% 		% \toprule
% 		\multirow{2}{1.5cm}{} & \multicolumn{2}{c}{ShapeNet} & \multicolumn{2}{c}{S3DIS} \\
% 		                        \cmidrule(l{2pt}r{2pt}){2-3} \cmidrule(l{2pt}r{2pt}){4-5}
% 		         & $d_J$ & F1 & $d_J$ & F1 \\ \midrule
% 		% Cylinder & 0.28 & 0.84 & 0.61 & 0.57 \\
% 		LassoNet & 0.214 & 0.873 & 0.195 & 0.886 \\ 
% 		\bottomrule
% 	\end{tabular}
% 	\vspace{2mm}
% 	\caption{Performance of a single modeled trained on both ShapeNet and S3DIS annotations.}
% 	\vspace{-4mm}
% \end{table}

We also tried to train a single model for both ShapeNet and S3DIS annotations, yielding an average $d_J$ of 0.214 and $F1$ score of 0.873.
The performance drops in comparison with those achieved by LassoNet using separate models (see Tab. 2).
The cause for performance dropping mainly comes from large differences between ShapeNet and S3DIS annotations:
1) numbers of points in S3DIS point clouds are almost a hundred times greater than those in ShapeNet point clouds, 
2) point densities in the two corpora are very different, as S3DIS point clouds are collected by sensing of real-world indoor rooms, while ShapeNet point clouds are samples from synthetic CAD models,
and 3) selections for S3DIS point clouds are usually smaller regions, in comparison to those for ShapeNet point clouds. 
Nevertheless, the results still outperform basic CylinderSelection.

% \cmo{
% using the hyper parameters for S3DIS annotations,
% since the number of points of S3DIS point clouds is larger.
% Overall, 
% although the model 
% performs worse than the models trained separately on each corpus in Sec.~\ref{ssec:eval},
% it still achieves a competitive Jaccard Distance and $F1$ score 
% compared to the CylinderSelection.
% We suspect
% the performance drops mainly due to the large difference between ShapeNet and S3DIS,
% not only because the number of points of S3DIS is almost a hundred times greater than that of ShapeNet, 
% but also because S3DIS and ShapeNet come from different domains (\emph{i.e.}, S3DIS from real-world indoor areas scanning
% while ShapeNet from synthetic CAD models).
% Thus, it is challenging to use a single model to fit them at once.
% }

Training a single model for multiple datasets is a challenging task.
As for now, employing different parameter settings for different datasets is practically more feasible.
Many recent deep neural network models for point cloud processing, such as MCCNN~\cite{hermosilla2018mccnn}, also trained separate models for different datasets.
A potential solution is domain adaptation~\cite{ben2010theory}, especially multi-source domain adaptation, 
that has proven beneficial for learning source data from different domains~\cite{Hajiramezanali_2018_domaina}.
Furthermore, domain adaptation can also 
learn a well-performing model 
for target data that exhibit different but related distributions with source data,
% from source data distributions 
% on a different but related target data distribution,
thereby improving generalizability of the model.
Thus, we consider domain adaptation as an important direction for future work.

% \cmo{
% Nevertheless, our method still has a good generalizability and robustness on the point clouds in the same domain.
% First,
% % in our experiments,
% the density of point clouds is diverse,
% ranging from 10K to 6M $points/m^3$ in ShapeNet 
% and 0.2K to 60K $points/m^3$ in S3DIS,
% as well as the number of points (as mentioned in Sec.~\ref{ssec:data_pre}). 
% Second,
% the intention filtering, 
% which filters the points based on the lasso drawn by the user, 
% changes both the number of points and volume of each point cloud, 
% leading to more diverse densities
% and number of points of the datasets.
% Third,
% our model does not require to be trained on each point cloud.
% In our model experiments, 
% all the point clouds in testing set had not been trained on the models, 
% which means that our models trained on some point clouds
% achieve good performance on other point clouds with varying densities and number of points.
% }

%%%%%%%%%%%%%%%%%%%%%%%%%%%%%%%
\subsection{Limitations and Future Work}

Though the model experiment and user study have demonstrated that LassoNet advances prior methods, there are several limitations.

\begin{itemize}[leftmargin=*]
\item
All deep neural network models, no matter supervised or unsupervised, require tremendous amounts of training data to generate high prediction accuracy.
We tackle this issue using a new dataset with over 30K records generated by professional annotators.
\cmo{It is also feasible to extend this dataset by synthesizing variations from existing records~\cite{Fan2018}}.
Yet, there can still be certain scenarios not covered by the training data.
% On this point, CylinderSelection and CAST show more flexibility, since they do not require training. 
% Thus, for unknown datasets, which have not been studied and require to be examined more closely to discover unexpected patterns, CAST or CylinderSelection are more appropriate.

\item
When making refinements, users would like to select only a few points.
As observed by the participants, 
LassoNet tends to expand the selection to some closely correlated points.
A feasible solution here would be to add a conditional statement in LassoNet: 
when naive selection detects only a few points being selected, 
LassoNet automatically returns these points as output.

\end{itemize}

\noindent We also identify several promising direction for future work:
\begin{itemize}[leftmargin=*]

\item 
A first and foremost work is to update our backbone network to state-of-the-art deep learning models for processing point clouds.
% During the period of this study, 
% networks that are more advanced than PN++ have been proposed.
For example, Hermosilla et al. proposed MCCNN that utilizes Monte Carlo up- and down-sampling to preserve the original sample density, making it more suitable for non-uniformly distributed point clouds~\cite{hermosilla2018mccnn}.
We consider MCCNN as an important future improvement to enhance the \emph{effectiveness} and \emph{robustness} of LassoNet.

\item 
Second, we would like to develop visual analytics to get a better understanding of what has the network learned,
which currently
%Currently the learning process 
is a `blackbox'.
As for now, we suspect that the network has modeled several intrinsic properties of point clouds, 
including i) local point density, as astronomic point clouds exhibit;
ii) symmetric property, as indicated by airplane wings;
iii) heat kernel signature~\cite{sun_2009_concise, bronstein_2010_scale}, 
as the network can 
%correctly 
segment seat and legs of a chair (Fig.~\ref{fig:results_example1} (left)).
The issue calls for more visual analytics to `\textit{open the black box}'~\cite{Liu2017, Hohman2018}.

\item 
Last but not least,
we would like to incorporate more parameters in the mapping function, 
such as FOV and stereoscopic projection.
Up to this point, we have only modeled viewpoint in terms of camera position and direction,
but not other parameters.
Modeling these parameters would be necessary and interesting, 
% as the settings may greatly affect how users perceive point clouds.
% For instance, projecting point clouds from different angles or using different projection types 
% can impose great effects of the density on 2D surface.
% Doing this would be interesting, 
as it can potentially extend the applicability of LassoNet to many other scenarios, e.g., to improve selection in VR/AR environments (e.g.,~\cite{tong_2017_glyphlens, hurter_2018_fiberclay}).

\end{itemize}

% Last but not least, we believe LassoNet can be applied in many applications, such as to facilitate interactions with volume visualizations that represent scalar volume data organized in 3D grids~\cite{Elvins:1992:SAV:142413.142427}.

\if 0
\yu{for now, I can only come up with these. will add more later}

1. From generalizability point of view:

CAST, the family density-based selection techniques, is  suitable for selecting 3D point clouds with density variations. With the input lasso users are able to select the intended 3D point cloud cluster that can be recognized on 2D screen.

However, in case there is no such information in the datasets, then CAST is working as CylinderSelection. 

On this point, LassoNet shows better generalizability. From the user study we can see that D1 and D2... and D3, the results are still as good as CAST.

2. From data scenario point of view

CAST can be used to select unknown data, for instance, in exploring an astronomical data which may have no prior information. 

But LassoNet, as a deep learning technique, needs to be trained.

3. whether we can combine these two methods (or some of the features) in the future?
\fi
\section{Conclusion}

We presented LassoNet, a new learning-based approach of lasso-selection for 3D point clouds built upon deep learning.
Our approach can be readily applicable to any scenario where one has a set of unordered points ($P$), a 2D surface for visualizing the points ($V$), and a lasso on the surface ($L$).
Essentially, LassoNet can be regarded as an optimization process of finding a functional latent mapping function $f(P, V, L): \rightarrow P_s$ such that $P_s$ matches best with a user's intention of selection $P_t$.
To learn such an optimal mapping, we created a new dataset with over 30K selection records on two distinct point cloud corpora.
LassoNet also integrates a series of dedicated modules including coordinate transformation, intention filtering, and furthest point sampling to tackle the challenges of data heterogeneity and scalability.
A quantitative comparison with two prior methods demonstrated robustness of LassoNet over various combinations of 3D point clouds, viewpoints, and lassos in terms of effectiveness and efficiency.

\acknowledgments{
The authors wish to thank the anonymous reviewers for their valuable comments.
This work was supported in part by
National Natural Science Foundation of China (No.61802388 and No.61602139).
This work is partially supported by a grant from MSRA (code: MRA19EG02).
}

\bibliographystyle{abbrv}

\bibliography{reference}
\end{document}